\documentclass[%
 reprint,
amsmath,amssymb,
aps,
pra,
floatfix,
]{revtex4-2}

\usepackage{placeins}
\usepackage{graphicx}%
\usepackage{dcolumn}%
\usepackage{bm}%
\usepackage{graphicx} %
\usepackage{dcolumn} %
\usepackage{bm} %
\usepackage{amsmath, amssymb, mathtools} %
\usepackage[caption=false]{subfig}
\usepackage{algorithmic} %
\usepackage{textcomp} %
\usepackage{xcolor} %
\usepackage{svg} %
\usepackage{pgfplots} %
\usepackage[percent]{overpic} %

\usepackage{physics}
\usepackage{soul}
\usepackage{amsfonts}
\usepackage{mathrsfs}
\usepackage{float}
\usepackage[colorlinks=true, linkcolor=blue, linkcolor=blue, citecolor=blue, urlcolor=blue]{hyperref}

\begin{document}

\preprint{APS/123-QED}

\title{Responsivity and Stability of Nonlinear Exceptional Point Lasers with Saturable Gain and Loss}%

\author{Todd Darcie}
 \affiliation{Department of Electrical and Computer Engineering, \\University of Toronto}%
 \email{todd.darcie@mail.utoronto.ca}
\author{J. Stewart Aitchison}%

\affiliation{%
  Department of Electrical and Computer Engineering, \\University of Toronto}

\date{\today}%

\begin{abstract}
The responsivity of perturbation sensing can be effectively enhanced by using higher-order exceptional points (HOEPs) due to their nonlinear response to frequency perturbations. However, experimental realization can be difficult due to the stringent parameter conditions associated with these points. In this work, we study an EP laser composed of two coupled nonlinear resonators that uses nonlinearity to simplify these tuning requirements. This system demonstrates a distinct cube-root response in the steady-state lasing frequency, with a constant of proportionality that depends on the distribution of linear and saturable gain and loss. This design freedom enables several orders of magnitude higher responsivity than systems with a single nonlinear resonator, which have been previously explored. Maximizing responsivity also improves the robustness of sensing performance against parametric errors. These features are derived from coupled mode theory and further supported by steady-state ab initio laser theory (SALT) results at several nonlinear EPs. Through linear stability analysis, we also identify regions of instability within the class-A regime that arise due to mode competition, which can be induced by asymmetric passive losses. In the class-B regime, we show that the interplay between gain dynamics and detuning can lead to restabilization at slow relaxation rates or higher inter-resonator coupling rates. This regime could be used to increase the maximum achievable responsivity of the system.

\end{abstract}

\maketitle

\section{\label{sec:intro} Introduction}

Exceptional points (EPs), which arise in non-Hermitian systems, are singularities where two or more eigenvectors (or eigenstates) coalesce \cite{Feng2017, ElGanainy2018, Miri2019}. The corresponding reduction in dimensionality makes these points extremely sensitive to external perturbations \cite{Wiersig2014,Wiersig2016,Wiersig2020}. While the response to perturbation about a diabolic point is linear \cite{Li2014}, the eigenvalue splitting around EPs is nonlinear, instead following an $n$th-root dependence \cite{Kato2012}.

While there have been several experimental demonstrations of sensing at second-order \cite{Hodaei2016,Chen2017,Zhao2018,Hokmabadi2019,Lai2019,Kononchuk2022,Zhang2024} and third-order \cite{Hodaei2017,Zeng2019} EPs that are isolated in parameter space, this approach comes with tight fabrication and/or tuning requirements. To reach a linear EP3, for instance, requires tuning six parameters \cite{Hodaei2017,Bai2022}.

The exceptional surface (ES), a two-dimensional surface of EPs, was introduced as a way to address this difficulty \cite{Zhong2019, Zhou2019}. Second-order exceptional surfaces (ES2s) have been demonstrated experimentally \cite{zhang2019,qin2021}, and other schemes have been proposed to extend the response to higher orders \cite{Yang2021,Kullig2023}. Rather than being PT-symmetric (involving asymmetric losses), this class of EPs originates due to asymmetric backscattering, which is perturbed by the presence of foreign nanoparticles. Extending this concept to achieve an enhanced response to frequency perturbation is not straightforward.

Another approach to reaching an EP is to exploit the nonlinearity inherent in systems with gain (due to saturation \cite{Hassan2015, ge2016, Bai2022, Ji2023, Darcie2023, Darcie2024} or the Kerr effect \cite{Chen2022}, for instance). Recent work has shown that a PT-symmetric EP3 can be reached in a binary system consisting of a nonlinear resonator with gain and a linear resonator with net loss \cite{Bai2022}. Reaching this EP reduces the number of tuning parameters to two, down from six for a linear EP3. Furthermore, this idea of exploiting nonlinearity can be extended to higher orders to yield an $n$th-root response in fewer than $n$ resonators via the introduction of ``hidden" dimensions \cite{Bai2023}.

In this work, we further extend this concept to the more general case where both resonators in a binary system are subject to varying amounts of saturable gain or saturable absorption. This gain is offset by passive losses, which are tailored such that a PT-symmetric EP is located at or above threshold. Unlike previous work looking at sensing with only a single nonlinear resonator (e.g., \cite{Bai2022}), this corresponds to the scenario where two identical resonators are fabricated using the same gain medium, then subjected to different amounts of pumping. Using coupled mode theory, we show that the frequency of the lasing mode still follows a cube-root response to frequency perturbation. However, the constant of proportionality can be several orders of magnitude higher than that in the system with only one nonlinear resonator. The upper limit of this signal scale factor is determined by either the minimum achievable passive losses or the maximum available gain. We also show that maximizing this scale factor makes the responsivity of the system much more robust to parametric errors (in gain or coupling rate, for instance), which are unavoidable when dealing with EPs located in isolated parameter spaces.

Unlike previous studies (e.g., \cite{Hodaei2016}), we consider the passive losses to be completely independent of one another. This treatment accounts for asymmetry in fabrication, as well as the possibility of output coupling from only one resonator. We find that a large contrast in passive losses can introduce instability in the EP mode due to the existence of broken-symmetry modes that can become energetically favored. Below this critical contrast, we confirm that the mode can still be unstable at the EP if the gain relaxation rate of the medium is too low. Expanding on these results, we then show that introducing a detuning between the resonators will eventually re-stabilize the system, effectively introducing a lower detection limit when operating in this regime.

To further investigate whether the cube-root response can be achieved in a physical system, we then study a laser consisting of coupled, differentially-pumped slabs using Steady-state Ab-initio Laser Theory (SALT) \cite{Esterhazy2014}. This technique allows the frequency of the lasing modes to be tracked without many of the usual assumptions used with coupled mode theory. We find that operating at a nonlinear EP can still be advantageous, but that an additional geometric tuning parameter is required if the spatial gain/mode profiles are inhomogeneous in each resonator. Furthermore, we find some qualitative agreement between SALT and CMT results for the signal scale factor of the system. 

\section{Coupled mode analysis}

\begin{figure}[ht] 
    \subfloat{%
        \includegraphics[width=.9\columnwidth]{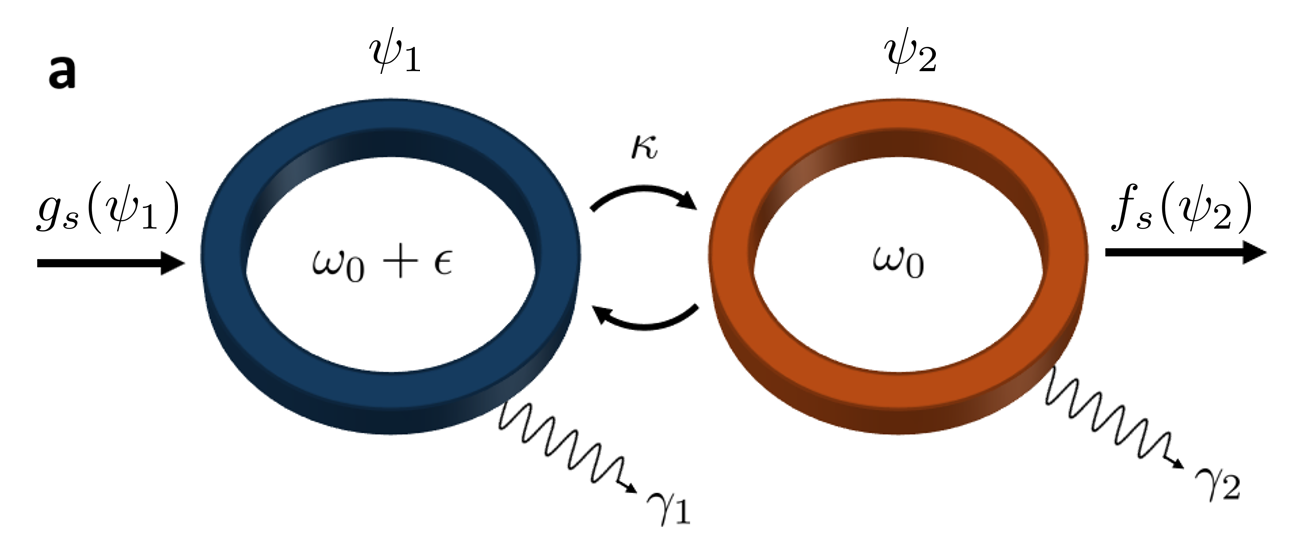}
    }%
    \\
    \subfloat{%
        \includegraphics[width=0.48\columnwidth, trim={.4cm .3cm .6cm .6cm}, clip]{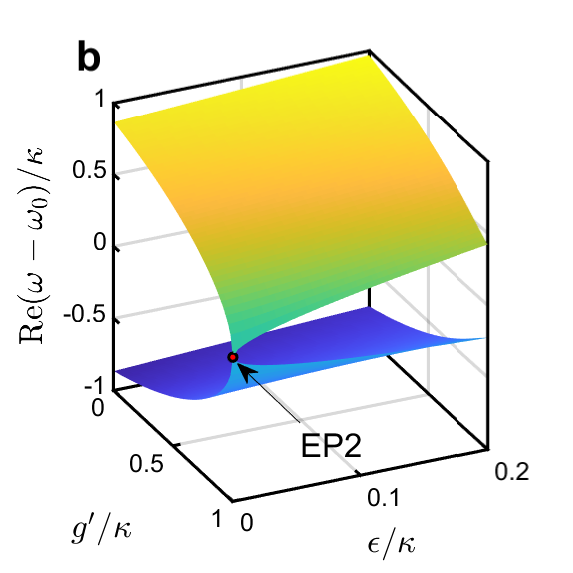}
    }%
    \hfill
    \subfloat{%
        \includegraphics[width=0.48\columnwidth, trim={.4cm .3cm .6cm .6cm}, clip]{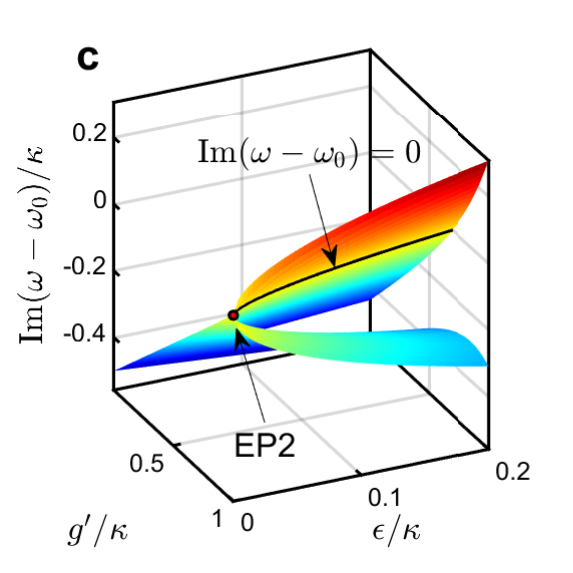}
    }
    \caption{(a) Schematic of the coupled resonator system under analysis.
    Resonator 1 experiences gain through external pumping and dissipation through scattering and output coupling. Resonator 2 experiences dissipation through absorption and scattering. (b) Real part of the eigenvalues $\omega$ of the binary coupled resonator system with loss fixed at $\gamma_2 = \kappa$ and $f_0 = 0$, as a function of gain $g'$ and detuning $\epsilon$. (c) Imaginary part of the eigenvalues.}
    \label{fig:eigs_binary}
\end{figure}

To model the dynamics of the binary system of conservatively coupled resonators, we employ a similar coupled mode formalism as in Refs.~\cite{Liu2017, Hodaei2016, Hodaei2017}. This approach is generally applicable to microdisks, microrings, or waveguides. A schematic of the system is shown in Fig.~\ref{fig:eigs_binary}(a).

The system is described by the Hamiltonian
\begin{equation} \label{H}
     H (\epsilon) = 
     \begin{bmatrix}
        \omega_0 + \epsilon + i(g_s - \gamma_1) &  \kappa\\
        \kappa & \omega_0 - i(f_s + \gamma_2)
    \end{bmatrix},
\end{equation}
which acts on the wave functions $|\psi\rangle = \left[\psi_1, \psi_2\right]^T$ in resonators 1 and 2. Here, $T$ denotes the matrix transpose. The intensity in resonator $j$ is $I_j = |\psi_j|^2$, which is normalized with respect to the saturation intensity and is therefore dimensionless. In addition, $\omega_0$ is the resonance frequency of each resonator in the absence of a frequency perturbation $\epsilon$, $\gamma_1$ and $\gamma_2$ are the passive decay rates of each resonator (due to scattering, output coupling, and radiation losses), and $g_s$ and $f_s$ are the saturated gain in resonator 1 and saturated absorption in resonator 2, respectively. Lastly, $\kappa$ is the coupling strength, which we assume to be real. The general steady-state saturation equations for the homogeneously broadened gain and absorption are
\begin{equation} \label{g_s}
    g_s = \frac{g_0}{1 + |\psi_1|^2},
\end{equation}
\begin{equation} \label{f_s}
    f_s = \frac{f_0}{1 + |\psi_2|^2},
\end{equation}
where $g_0$ and $f_0$ are the small-signal gain and loss, respectively.

The time evolution of the system is described by the Schrödinger-like equation
\begin{equation} \label{shrodinger}
    i \frac{\mathrm{d}}{\mathrm{d} t}|\psi\rangle = H(\epsilon)|\psi\rangle.
\end{equation}

In terms of $g_s$ and $f_s$, the characteristic equation of $H(\epsilon)$ can be expanded as
\begin{multline} \label{chareqn}
        (\omega - \omega_0)^2 + i(f' - g') (\omega - \omega_0) -  i\epsilon f' - \epsilon (\omega - \omega_0) \\
        + g' f' - \kappa^2 = 0,
\end{multline}
where $g' \equiv g_s - \gamma_1$ and $f' \equiv f_s + \gamma_2$ are the net modal gain and loss in resonators 1 and 2, respectively. From Eq.~\ref{chareqn}, we obtain the eigenvalues and eigenvectors:
\begin{equation} \label{vals}
    \omega_{1,2} =  \omega_0 + \frac{i(g' - f')}{2} + \frac{\epsilon}{2} \pm \frac{i \sqrt{(\Gamma - i\epsilon)^2 -4 \kappa^2}}{2},
\end{equation}
\begin{equation} \label{vects}
     |\phi_{1,2}\rangle =   
     \begin{bmatrix}
         i(g' + f') + \epsilon \pm i \sqrt{(\Gamma - i\epsilon)^2 -4 \kappa^2} \\
        2 \kappa
    \end{bmatrix},
\end{equation}
where $\Gamma \equiv g' + f'$ is the gain-loss contrast.

The thresholds of similar systems at zero detuning have been previously studied in Refs.~\cite{Hodaei2017, benzaouia2022}. The lasing solutions above threshold generally follow two types. First, there are PT-symmetric solutions, which arise when $|\Gamma| < 2 \kappa$. These modes are detuned in either direction from $\omega_0$ but have equal modal gain and symmetric intensity profiles. Lasing requires these modes to have real eigenvalues, which is achieved when $g' - f' = 0$, the condition for PT symmetry. Note that this condition can still be reached if $f_0 < 0$, meaning that both resonators have saturable gain, by offsetting the gain with passive losses.

There are also PT-broken modes, which arise when $|\Gamma| > 2 \kappa$ and are characterized by degenerate frequency but asymmetric intensity and modal gain. The lasing condition for PT-broken modes imposes that $g' f' = \kappa^2$.

The eigenvalues and eigenvectors are degenerate when $\epsilon = 0$ and $\Gamma = \pm 2 \kappa$, where $\Gamma = -2 \kappa$ is simply a mirrored version of the system with $\Gamma = 2 \kappa$.

Combining the conditions for PT-symmetric and PT-broken lasing modes gives $g' = f' = \pm \kappa$, which is the well-known condition for the PT-symmetric EP \cite{Feng2017}, residing at the transition between the two phases.

This transition is accessible above threshold ($I_{1,2} \geq 0$) at zero, one, or two specific values for the applied gain:
\begin{equation} \label{g_c}
g_c =
\begin{cases}
f_0 \dfrac{ \kappa + \gamma_1}{\kappa - \gamma_2}, 
& \text{if } (\gamma_2 \leq \kappa \ \text{and} \ f_0 + \gamma_2 \geq \kappa) \\
& \quad \text{or} \ (\gamma_2 > \kappa \ \text{and} \ f_0 + \gamma_2 < \kappa), \\
f_0 \dfrac{ \kappa - \gamma_1}{\kappa + \gamma_2}, 
& \text{if } f_0 + \gamma_2 \leq -\kappa.
\end{cases}
\end{equation}

The first expression is found by setting $I_{1} = I_2$, $g' = f' = \kappa$ and solving for $g_0$. 
The second expression arises from the fact that the roles of resonators 1 and 2 can be swapped if both resonators contain saturable gain ($f_0 < 0$). This corresponds to the EP with $\Gamma = -2\kappa$, so the analysis in the following section will require that the system be mirrored to analyze the frequency response around this EP.

\begin{figure}[ht]
    \centering
    \subfloat{%
        \includegraphics[width=.4\textwidth]{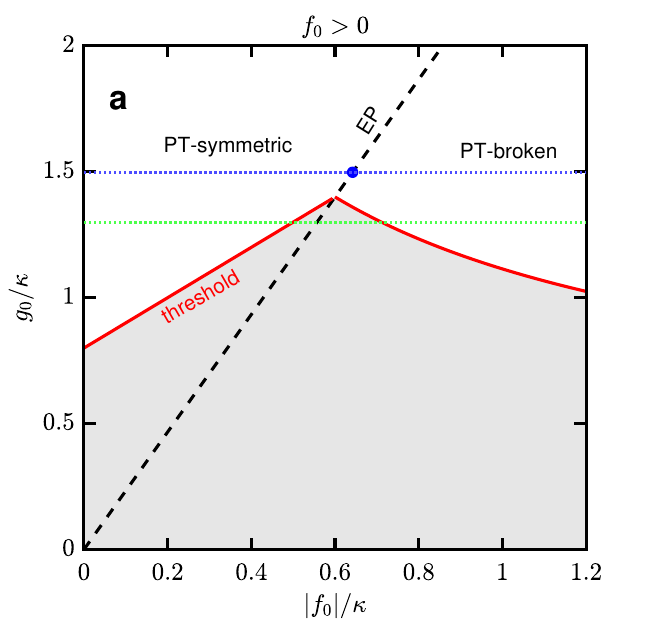}
    }
    \hfill
    \subfloat{%
        \includegraphics[width=.4\textwidth]{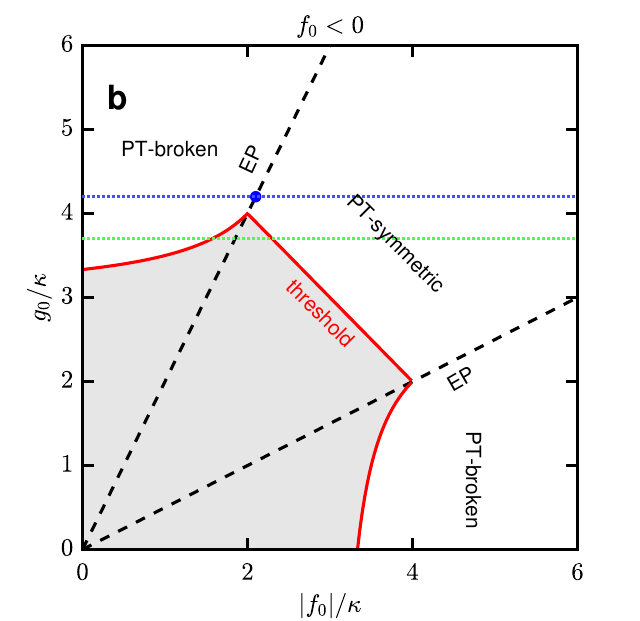}
    }
    
    \captionsetup{justification=justified, singlelinecheck=false, width=\linewidth}
    \caption{Phase map of the coupled resonator system. (a) Resonator 1 exhibits saturable gain while resonator 2 has saturable absorption, with passive losses $\gamma_1 = \gamma_2 = 0.4 \kappa$. The exceptional point (EP) is reached either by lasing into the PT-symmetric regime well above threshold and then increasing the absorption (blue dotted line), or by increasing the gain while keeping the absorption constant and greater than $\kappa - \gamma_2$. Along the green line, increasing $|f_0|$ results in loss-induced suppression and revival of lasing, as reported in \cite{peng2014}. (b) Both resonators possess saturable gain with passive losses $\gamma_1 = \gamma_2 = 3\kappa$. In this scenario, increasing $|f_0|$ along the green line leads to gain-induced suppression and revival of lasing.}
    \label{fig_threshold_map}
\end{figure}

Figure~\ref{fig_threshold_map}(a) and (b) map these thresholds and phase transitions for $f_0 > 0$ and $f_0 < 0$, respectively. For (a), the EP can be reached either by first increasing $g_0$ until lasing begins with unbroken symmetry, and increasing $f_0$ (blue dotted line). This approach is used for a similar system in Ref.~\cite{Hodaei2016}. However, this approach requires tuning of both absorption and gain, and there is the possibility of not reaching the EP at all, and instead following the path of loss-induced suppression and revival of lasing (greed dotted line). Another approach is to start with $f_0 + \gamma_2 > \kappa$ and then increase $g_0$ until lasing begins with broken symmetry, and then continue to increase $g_0$ until the EP is reached. The EP can be reached via a similar process for $f_0 < 0$, which is shown in Fig.~\ref{fig_threshold_map}(b).

In parameter regions where an EP exists at an applied gain of $g_c$ (with a modal gain of $g' = \kappa$), there can also be other lasing modes with the same applied gain but different saturated gain and modal gain. This occurs because the modal gain is not directly controlled. These other modes have broken symmetry and emerge when the radicand of Eq.~\ref{vals} is positive but is compensated by $g' \leq f'$ such that the eigenvalue remains real. The derivation of these additional modes is shown in Appendix~\ref{non_degenerate_modes}.

The system will only remain at the EP if the lowest modal gain $g'$ of all other physical modes is greater than $\kappa$. Otherwise, the EP mode is unstable, and the system will evolve into a PT-broken lasing mode due to gain competition (see Appendix~\ref{non_degenerate_modes}).

When the gain and loss are considered as linear, the frequency splitting from the EP in response to a small perturbation $\epsilon > 0$ is $\Re(\omega_1 - \omega_2) \approx \sqrt{2 \kappa \epsilon}$. This matches the previous results of Refs.~\cite{Hodaei2017}. However, the perturbation also affects the imaginary parts of $\omega_1$ and $\omega_2$; $\Im(\omega_1) \approx \sqrt{2 \kappa \epsilon}/2$ corresponds to an additional amplification of mode 1, while $\Im(\omega_2) \approx -\sqrt{2 \kappa \epsilon}/2$ corresponds to exponential decay of mode 2. The former results in a decrease of $g_s$ according to Eq.~\ref{g_s}, pushing the system further from the EP. Since a steady-state lasing mode must have real eigenvalues \cite{Lu2015}, the amount of saturation will also depend on $\epsilon$. This is shown by the solid black line in Fig.~\ref{fig:eigs_binary}(b), which shoes that $g'$ remains less than $\kappa$ so as to ensure that $\Im{\omega-\omega_0} = 0$

\section{Frequency response with gain and loss saturation}\label{gainandloss}

In this section, we will analyze the frequency response to small detuning, assuming $\epsilon \ll \kappa$. To begin, assume that the small-signal gain and loss are tuned such that the system is initially at an EP at $\epsilon = 0$. Furthermore, we assume that this EP is located above threshold such that $g_0 = g_c$ and $g' = f' = \kappa$.

We then assume that the detuning results in saturation of the gain and absorption according to a characteristic leading exponent $u > 0$:
\begin{equation} \label{g_u}
    g' = \kappa - b_g \left|\epsilon\right|^u,
\end{equation}
\begin{equation} \label{f_u}
    f' = \kappa - b_f \left|\epsilon\right|^u.
\end{equation}

For real eigenvalues to exist, the imaginary part of Eq.~\ref{chareqn} must be zero, which yields $g' / f' = (\omega - \omega_0 - \epsilon)/(\omega - \omega_0)$. We note that this equation could not be satisfied for nonzero $\epsilon$ if $g'$ and $f'$ were to saturate with different leading exponents, which justifies the above assumption. Combining this with Eqs.~\ref{g_u} and \ref{f_u} and keeping only the lowest order terms in $\epsilon$ gives
\begin{equation} \label{charexp}
    \left|\epsilon\right|^u = \frac{\kappa \epsilon}{(\omega - \omega_0) (b_f - b_g)}.
\end{equation}

Substituting into Eq.~\ref{chareqn} gives
\begin{equation} \label{chareqn_real}
    (\omega-\omega_0)^3 - \epsilon (\omega - \omega_0)^2 - \kappa^2 \epsilon \frac{\bar{b}}{\delta} = 0,
\end{equation}
where $\bar{b} = (b_g + b_f)/2$ and $\delta = (b_g - b_f)/2$. Since the eigenvalues are near-degenerate at the EP, the second term is expected to be negligible compared to the third term.

We therefore see that $\Re(\omega-\omega_0)$ for the real root scales as
\begin{equation} \label{freq_response}
    \Re(\omega-\omega_0) \approx \sqrt[3]{\kappa^2 \epsilon \bar{b}/\delta}.
\end{equation}
This agrees with previous results demonstrating a scaling of $\sqrt[3]{\kappa^2 \epsilon}$ in a system with only one nonlinear resonator ($f_0 = 0$, $\gamma_2 = \kappa$) \cite{Bai2022}. Since the loss in resonator 2 is entirely passive, this case requires that $b_f = 0$ and $\bar{b}/\delta = 1$.

Applying the result in Eq.~\ref{freq_response} to Eq.~\ref{charexp} gives a characteristic exponent of $u = 2/3$. The equation then reduces to
\begin{equation} \label{b}
    \bar{b} = \frac{\kappa}{8 \delta^2}.
\end{equation}

The corresponding signal enhancement factor (SEF) for the more general case, as defined in \cite{Wang2020, Bai2022}, can be expressed as
\begin{equation} \label{SEF}
    SEF \equiv \left|\partial \Re(\omega-\omega_0) / \partial \epsilon\right|^2 = \frac{2}{9}\tilde{b}|\epsilon|^{-4/3},
\end{equation}
where $\tilde{b} = \bar{b} \kappa$ will be referred to as the signal scale factor. For comparison, the SEF for a linear second-order EP is $\kappa/8\epsilon$, which is obtained by taking the real part of Eq.~\ref{vals} with $g_s = g_0$ and taking lowest order terms.

A closed-form expression for $\tilde{b}$ is derived in Appendix~\ref{sens_derivation} by combining the eigenvector equations (Eq.~\ref{vects}) with the saturation equations (Eqs.~\ref{g_s} and \ref{f_s}). The result is
\begin{equation} \label{eq:tilde_b_updated}
    \tilde{b} = \dfrac{1}{2} \left( \dfrac{\kappa}{f_0 ( \gamma_1 + \gamma_2 )} \cdot T \right)^{2/3},
\end{equation}
where
\begin{equation} \label{eq:T_simplified}
    T = 2 (\kappa + \gamma_1)(\kappa - \gamma_2)^2 - f_0 \left[ \kappa ( \gamma_1 - \gamma_2 ) - 2 \gamma_1 \gamma_2 \right ].
\end{equation}

Note that $\bar{b}$ and $\tilde{b}$ can be normalized by $\kappa^{1/3}$ and $\kappa^{4/3}$, respectively, to express the signal scale factor in dimensionless form. 

\begin{figure}[ht]
    \centering
    \subfloat{%
        \includegraphics[width=.4\textwidth]{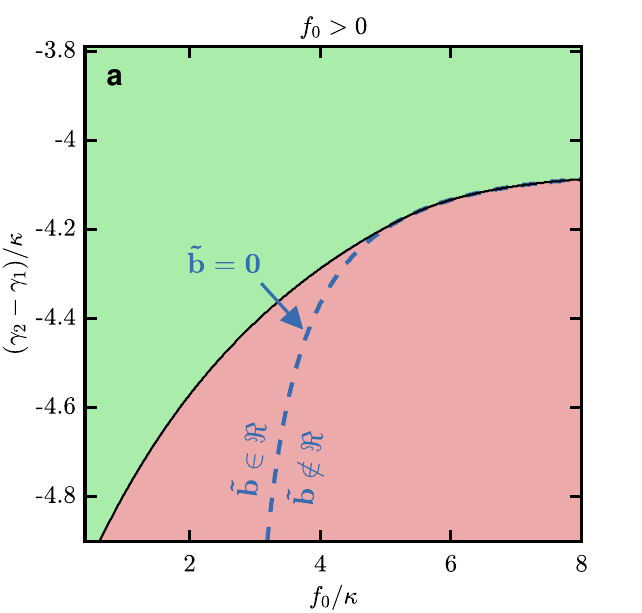}
    }
    \hfill
    \subfloat{%
        \includegraphics[width=.4\textwidth]{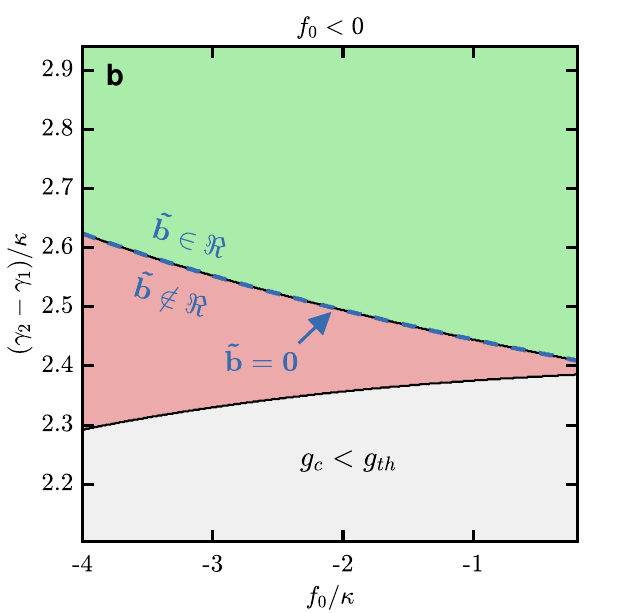}
    }
    
    \captionsetup{justification=justified, singlelinecheck=false, width=\linewidth}
    \caption{Stability map of the EP mode under variation of (a) absorption $f_0$ or (b) gain $-f_0$ and asymmetry in passive losses $\gamma_2 - \gamma_1$. The green regions represent parameter combinations when the EP mode is stable according to the modal analysis in Appendix~\ref{non_degenerate_modes}, which assumes fast inversion relaxation. Conversely, the red regions represent unstable combinations, and the light grey regions represent combinations where no EP appears above the lasing threshold. The solid black lines mark the borders between these regions. Blue dashed lines show where the signal scale factor $\tilde{b} = 0$, which is the border between real and complex values. The other parameters are $(\gamma_1 + \gamma_2)/2 = 1.5$ and $2.2$ for (a) and (b), respectively.}
    \label{b_stability_comparison}
\end{figure}

From Eq. \ref{eq:tilde_b_updated}, we find that $\tilde{b}$ can be either zero or complex for certain parameter sets. Specifically, $\tilde{b}$ is complex if $T$ and $f_0$ have opposite signs. This requires a large asymmetry in the passive losses of the two resonators. In Appendix~\ref{b_stability}, we show that these complex values correspond to parameter sets where the EP mode is unstable. For $f_0 < 0$, the inverse is also true; parameters resulting in real $\tilde{b}$ ensure the stability of the EP mode. This is shown in Fig.~\ref{b_stability_comparison}(b). While there is some correspondence between these conditions for $f_0 > 0$, Fig.~\ref{b_stability_comparison}(a) shows that the conditions for stability are slightly more restrictive than the conditions for real $\tilde{b}$.

For sensing, maximizing $\tilde{b}$ is of paramount interest. From Eq.~\ref{b}, it is clear that minimizing the passive losses is preferred. However, some passive loss is unavoidable with any experimental implementation (e.g., due to scattering and output coupling). Furthermore, the available gain is always limited to some maximum value. We therefore consider the task of finding the optimal $\kappa$ and $f_0$ for minimum achievable passive losses $\gamma_1, \gamma_2$ and a maximum gain $g_{max}$.

The conditions for the existence and stability of an EP above threshold constrain $\kappa$ and $f_0$ for these given gain and loss values. For an EP to be accessible also requires $g_c < g_{max}$, which provides a further constraint.

\begin{figure*}[ht]
    \centering
    \subfloat{%
        \includegraphics[trim=0cm 0cm 1cm .4cm, clip, width=0.3234\textwidth]{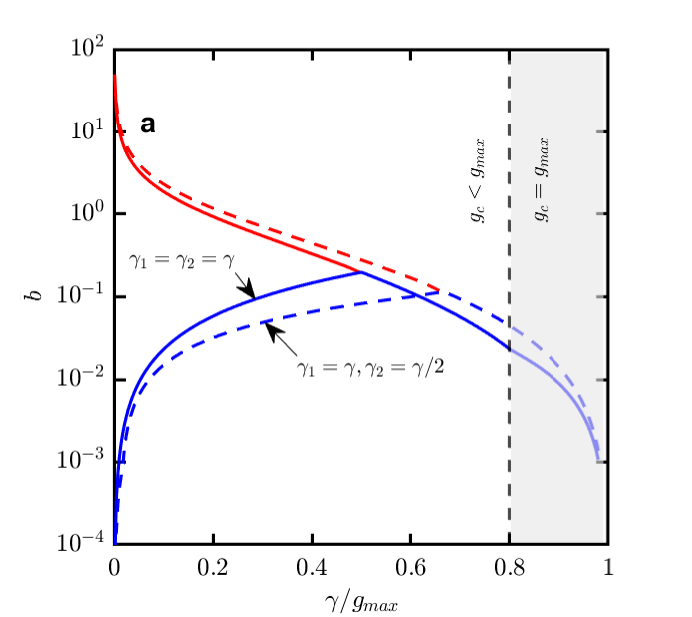}
    }%
    \hfill
    \subfloat{%
        \includegraphics[trim=0cm 0cm 1cm .4cm, clip, width=0.3234\textwidth]{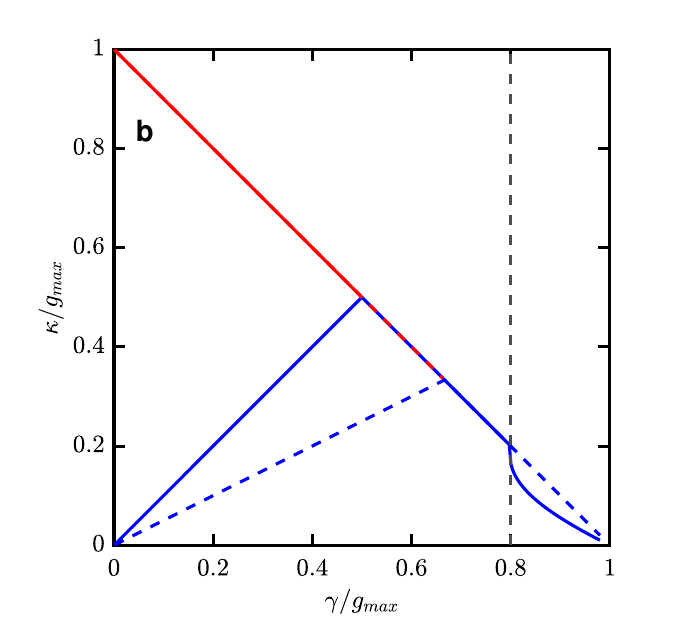}
    }%
    \hfill
    \subfloat{%
        \includegraphics[trim=0cm 0cm 1cm .4cm, clip, width=0.3234\textwidth]{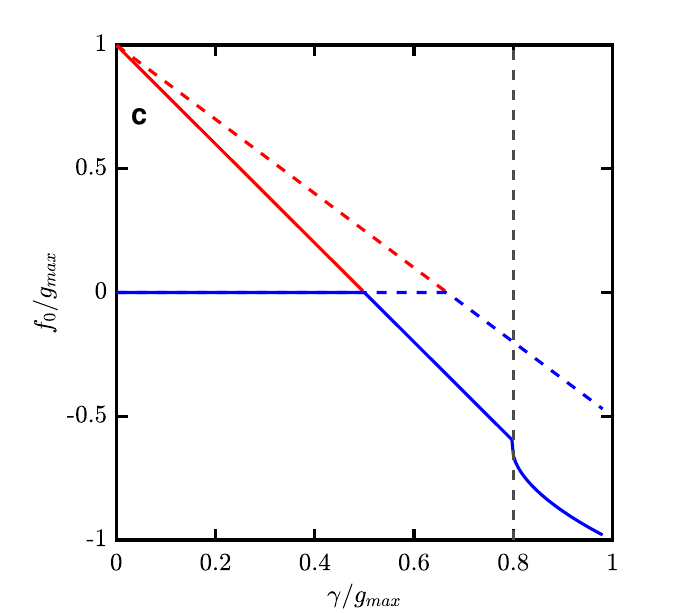}
    }%
   
     \caption{(a) Maximum signal scale factor $\tilde{b}$ as a function of passive loss and maximum applied gain $g_{max}$. The parameters of the system are subject to the constraints in Eq.~\ref{g_c}, which are required to lase at an EP at or above threshold. We also impose the constraint $g_c \leq g_{max}$, which holds with equality in the shaded region. The red and blue lines correspond to the solutions with saturable absorption ($f_0 > 0$) and saturable gain ($f_0<0$), respectively. (b) Optimal coupling rate $\kappa$ and (c) optimal absorption (gain) $f_0$ ($-f_0$) to achieve the maximum $\tilde{b}$ in (a).}
    \label{fig_sens}
\end{figure*}

Figure~\ref{fig_sens}(a) shows the maximum achievable signal scale factor for a given ratio of passive loss to maximum gain. We see that for a given passive loss, saturable absorption (red lines) is always preferable to saturable gain in both resonators (blue lines). These correspond to the first and second sets of conditions in Eq.~\ref{g_c}. This is true for both symmetric (solid lines) and asymmetric (dashed lines) loss distributions. For the later, we choose $\gamma_1 = 2 \gamma_2$ as an example to demonstrate a possible configuration where output coupling occurs from resonator 1 but not resonator 2. 

The values for $\kappa$ and $f_0$ where this maximum scale factor is reached are also shown in Fig.~\ref{fig_sens}(b) and (c), respectively. In (b), we observe that the key advantage of using saturable absorption is that the coupling rate can be increased at the same time as decreasing loss, affecting both the numerator and denominator of Eq.~\ref{b}. With $f_0 \leq 0$, on the other hand, the coupling rate must be decreased as the loss is decreased.

With both symmetric and asymmetric passive losses, there is a critical passive loss beyond which we require gain in both resonators for the EP to be accessible. Below this value, the saturable gain configuration is optimized for $f_0 \rightarrow 0^{-}$, which is equivalent to having a single nonlinear resonator coupled to a passive resonator. Below the critical passive loss, Fig.~\ref{fig_sens}(a) therefore provides a direct comparison between our system, with two nonlinear resonators, and the system studied in Ref.~\cite{Bai2022}. We see that the performance is equivalent at the critical passive loss, but that the signal scale factor can be improved by several orders of magnitude if the passive losses are low compared to the maximum gain.

\section{Impact of parametric errors}

The response obtained in the previous section relies on the system being biased exactly at the EP when the detuning between resonators is zero. This requires that the applied gain is exactly equal to $g_c$. In a practical experiment, the gain could change over the duration of a measurement due to thermal effects, electrical power fluctuations, mechanical instabilities, or other factors. Conversely, the inter-resonator coupling strength $\kappa$ could change. These two errors are equivalent in the sense that if the gain changes, there can still be a different value of $\kappa$ such that the new gain corresponds to $g_c$ at a new EP. We refer to this value as $\kappa_{EP}$, which is the coupling strength required to reach the EP for a given gain.

\begin{figure*}[ht]
    \centering
    \subfloat{%
        \includegraphics[trim=0cm 0cm 1cm .4cm, clip, width=0.3234\textwidth]{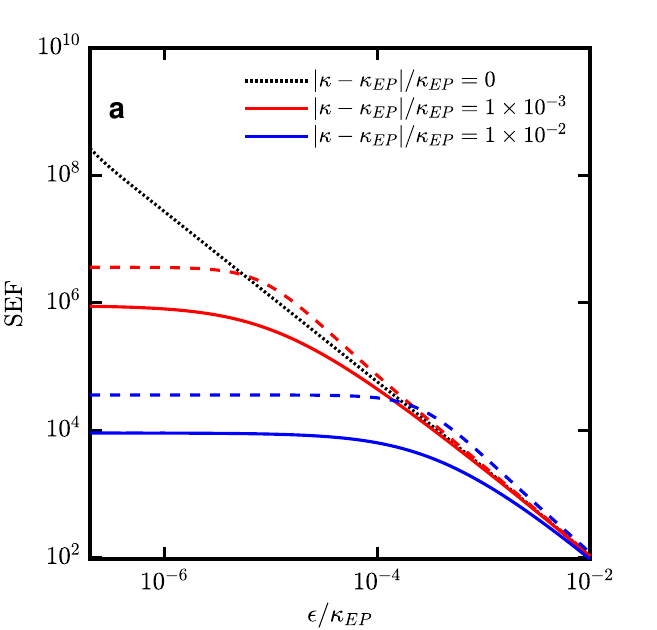}
    }%
    \hfill
    \subfloat{%
        \includegraphics[trim=0cm 0cm 1cm .4cm, clip, width=0.3234\textwidth]{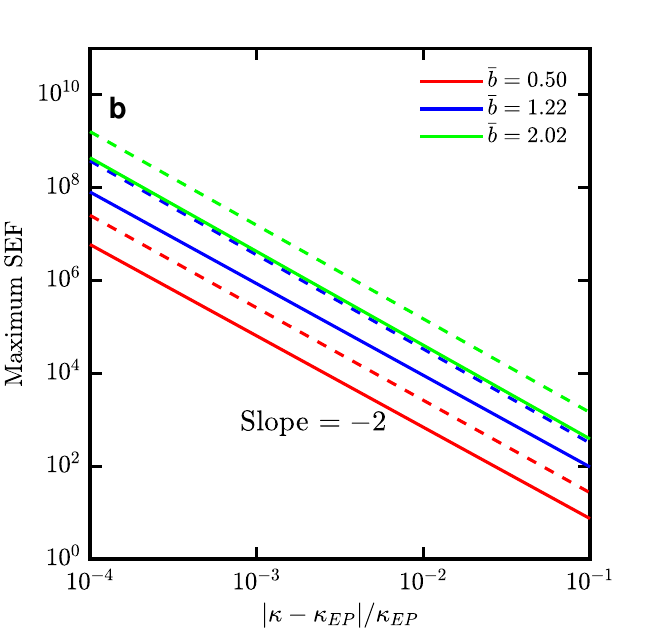}
    }%
    \hfill
    \subfloat{%
        \includegraphics[trim=0cm 0cm 1cm .4cm, clip, width=0.3234\textwidth]{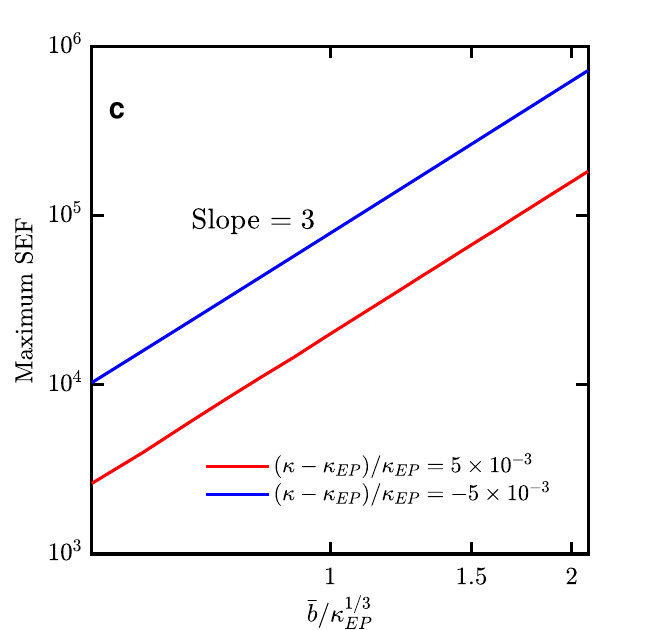}
    }%
   
     \caption{(a) Signal enhancement factor $SEF$ (Eq.~\ref{SEF}) as a function of detuning with coupling rate $\kappa$ positioned at $\kappa_{EP}$ (dotted line), slightly higher than $\kappa_{EP}$ (solid lines), and slightly lower than $\kappa_{EP}$ (dashed lines). $f_0$, $\gamma_1$, and $\gamma_2$ are chosen such that the signal scale factor $\bar{b}$ is equal to $1.22/\kappa_{EP}^{1/3}$. (b) The maximum achievable $SEF$ as a function of the offset between $\kappa$ and $\kappa_{EP}$. Different values of $\bar{b}$ are shown (which are normalized by $\kappa_{EP}^{1/3}$), with higher $\bar{b}$ resulting in greater maximum $SEF$ for a given coupling rate. (c) Maximum $SEF$ as a function of $\bar{b}$ at a constant coupling rate, which is slightly offset from the EP.}
    \label{fig_param_err}
\end{figure*}

Rather than continuing to increase with decreasing detuning, these parametric errors result in a plateau in the $SEF$ below a certain value of $\epsilon$, as shown in Fig.~\ref{fig_param_err}(a). In this region, the system behaves linearly since $\epsilon$ is insignificant compared to the parametric distance from the EP. The level of this plateau is higher if the coupling rate is slightly below its EP value, as opposed to slightly above (by the same amount). These two conditions correspond to the PT-broken and PT-symmetric regimes, respectively.

Notably, the responsivity advantage does not disappear in this region. While the frequency response changes from cube-root to linear, the constant of proportionality is still affected by the proximity to the EP.

Figure~\ref{fig_param_err}(b) shows the level of this plateau, which corresponds to the maximum achievable $SEF$, as a function of the relative error between $\kappa$ and $\kappa_{EP}$. We find that the maximum $SEF$ is inversely proportional to the square of this relative error, regardless of its direction. This is justified analytically in supplementary Section II.

Examining a constant relative error, Fig.~\ref{fig_param_err}(c) shows that it is preferable to have $\kappa < \kappa_{EP}$, as in Fig.~\ref{fig_param_err}(b). Furthermore, we find that the maximum $SEF$ is approximately proportional to the cube of $\bar{b}$, the signal scale factor. This is notable because the $SEF$ itself (when biased exactly at the EP) is only linearly proportional to $\bar{b}$ (Eq.~\ref{SEF}). Therefore, a primary advantage of designing the system to maximize this factor is that much greater responsivity can be realized in the presence of unavoidable parametric errors. 

\section{Stability analysis} \label{sec_stability_analysis}

The stability of the lasing mode(s) is of particular interest for sensing. Previous analyses of degenerate and nearly-degenerate lasing modes have shown that gain medium dynamics can affect the stability of the system \cite{Liu2017, Burkhardt2015, benzaouia2022}. In particular, Benzaouia and colleagues show that a laser biased at an EP can be stable given a sufficiently high inversion relaxation rate \cite{benzaouia2022}. They base their analysis on a simple system consisting of coupled slabs, which is studied by linearizing the Maxwell-Bloch equations. Other works have studied the stability of systems with exceptional points using a coupled-mode approach, without considering gain dynamics \cite{Zhou2016, Assawaworrarit2017, Bai2022}.

In the previous sections, the form of the nonlinear gain and absorption (Eqs.~\ref{g_s}--\ref{f_s}) assumes that the system is operating in the Class-A regime, where the polarization and inversion relaxation rates are much faster than the photon decay rate, such that the inversion and polarization can be adiabatically eliminated. In this regime, the system is generally stable unless there are highly asymmetric passive losses (Fig.~\ref{b_stability_comparison}). However, this Class-A assumption is easily violated in the case of strongly coupled microresonators (with high dissipation) or slow gain media. This includes most semiconductor lasers, as discussed in \cite{Yacomotti2023}. While the inclusion of these dynamics may not affect the eigenvalues of the lasing mode(s) or the steady-state saturation, they could affect the stability of the system.

In this work, we aim to retain the generality of the coupled-mode formalism while accounting for the dynamics of the inversion. To achieve this, we adopt dynamic equations for the saturated gain and absorption \cite{siegman1986}:
\begin{equation} \label{gaindynamics}
    \frac{\partial g_s}{\partial t} = -\gamma_{||} (1+|\psi_1|^2) g_s + \gamma_{||} g_0,
\end{equation}
\begin{equation} \label{lossdynamics}
    \frac{\partial f_s}{\partial t} = -\gamma_{||} (1+|\psi_2|^2) f_s + \gamma_{||} f_0,
\end{equation}
where $\gamma_{||}$ is the inversion relaxation rate. Here we have made the simplifying assumptions of small logarithmic gain and absorption ($g_s t_{rt},\ f_s t_{rt} \ll 1$), and we have ignored transverse variations of intensity.

We analyze the stability of the system for small perturbations about the steady-state lasing modes and calculate the Lyapunov exponents for the lasing mode as a function of detuning and gain relaxation rate. The details of this analysis are provided in Appendix~\ref{linear_stability_analysis}.

Lyapunov exponents $\Re(\sigma_j) > 0$ describe perturbations that grow exponentially in time, implying that the system is unstable. Conversely, if $\Re(\sigma_j) < 0$ for every $j$, the solution is stable against small perturbations.

\begin{figure*}[ht]
    \centering
    \begin{minipage}{0.48\textwidth}
        \subfloat{%
            \includegraphics[trim=.0cm 0cm 1cm .3cm, clip, width=0.98\linewidth]{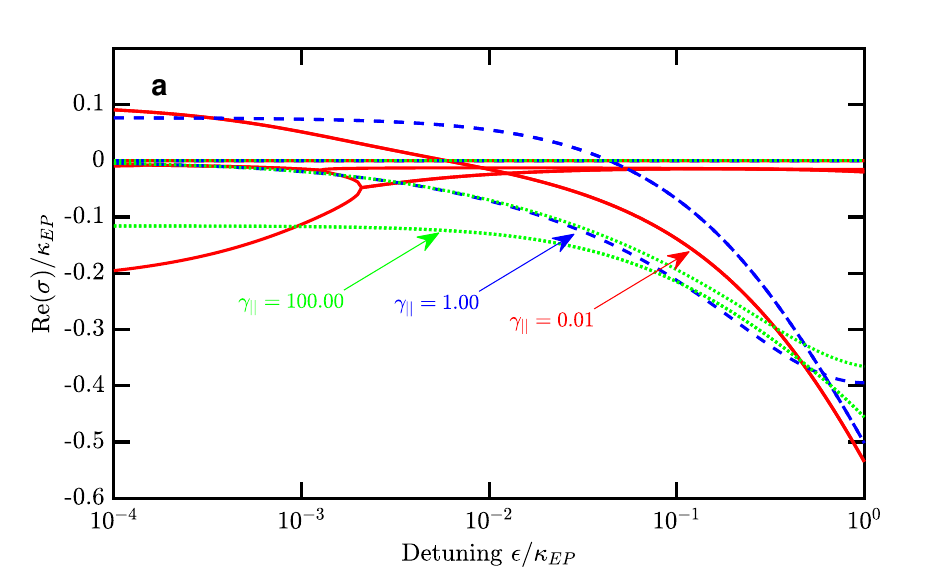}
        }
    \end{minipage}
    \begin{minipage}{0.48\textwidth}
        \subfloat{%
            \includegraphics[trim=.0cm 1.1cm 1cm .3cm, clip, width=0.98\linewidth]{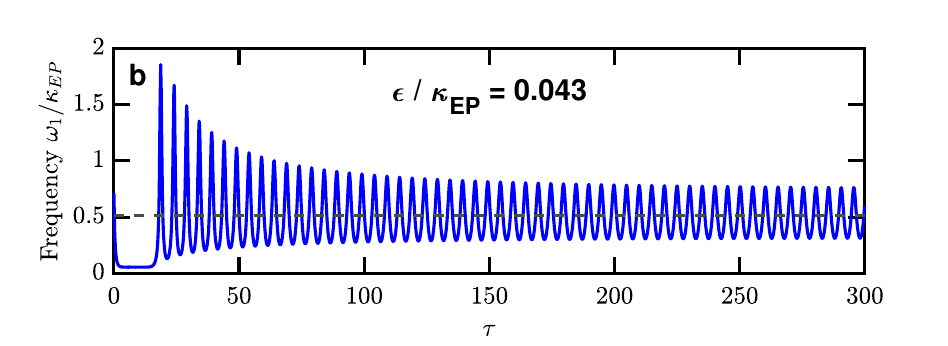}
        }\\
        \subfloat{%
            \includegraphics[trim=.0cm 0cm 1cm .3cm, clip, width=0.98\linewidth]{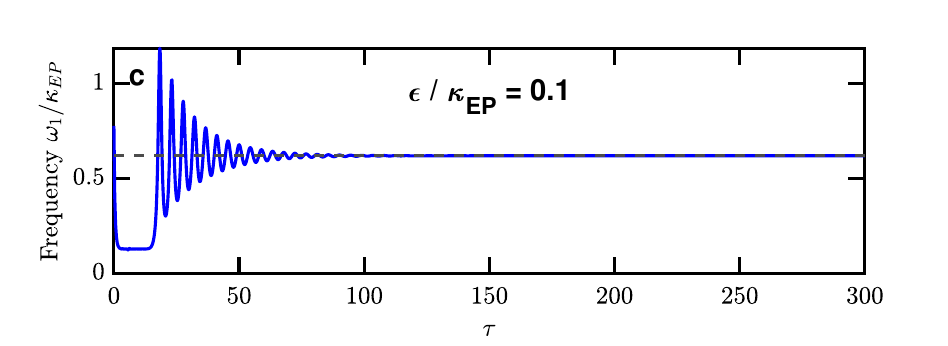}
        }
    \end{minipage}
    \\
    \subfloat{%
        \includegraphics[trim=.0cm 0cm 1cm .3cm, clip, width=0.48\textwidth]{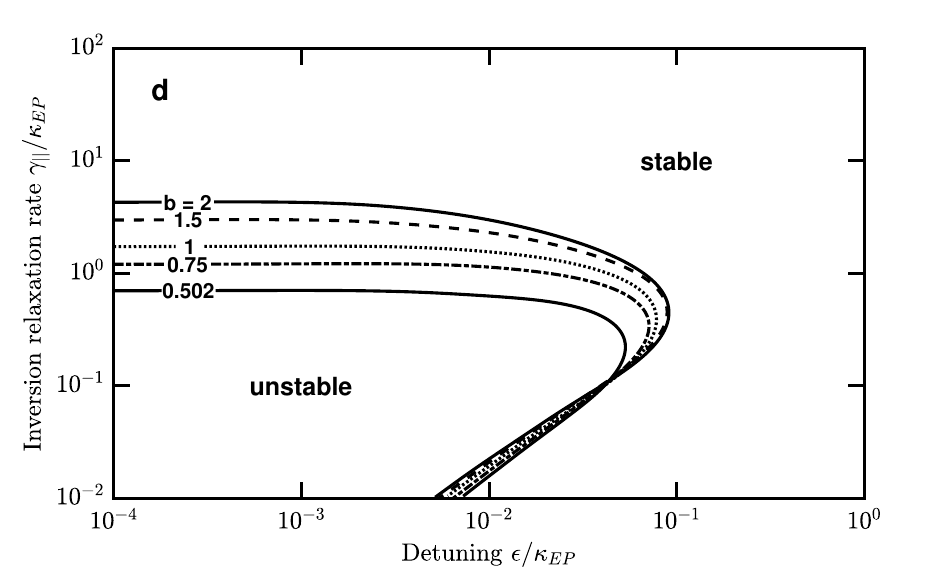}
    }%
    \subfloat{%
        \includegraphics[trim=.0cm 0cm 1cm .3cm, clip, width=0.48\textwidth]{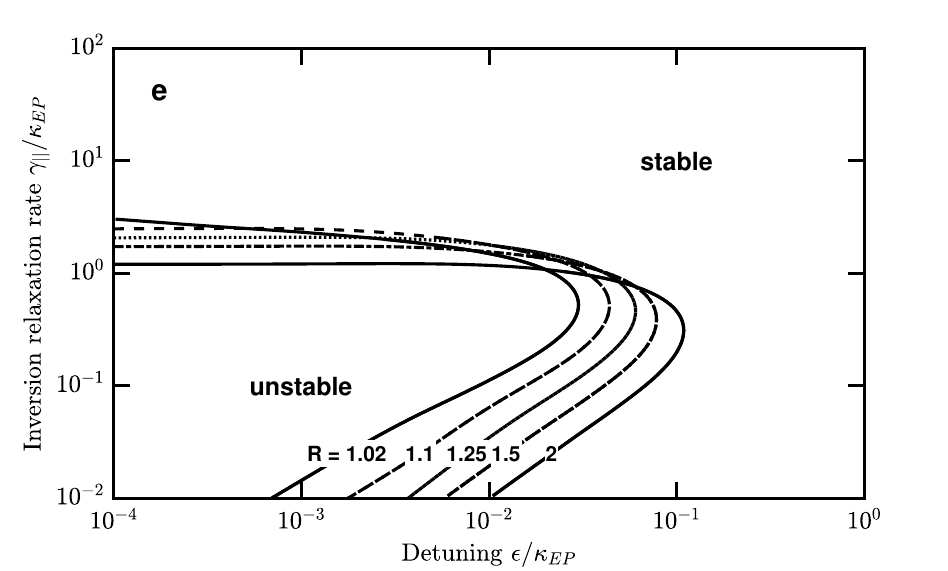}
    }

    \caption{(a) Lyapunov exponents $\Re(\sigma)$ for the lasing mode versus detuning $\epsilon$, normalized by coupling rate $\kappa$. Results are shown for $\gamma_{||} \in \{ 0.01, 1, 100 \}$. The other model parameters are chosen such that $\bar{b} = 1/\kappa^{1/3}$, $g_{\text{th}} = \kappa$, and $R = g_c/g_{\text{th}} = 1.5$. Some eigenvalues have a real part below the axis limits, so they are not shown. Time-domain fluctuations in lasing frequency for $\gamma_{||} = 1$ at $\epsilon/\kappa = 0.043$ (b) and $0.1$ (c). Boundaries between unstable and stable configurations for (d) different signal scale factors $\bar{b}$ (normalized by $\kappa^{1/3}$), with $g_{\text{th}} = \kappa$, $R = 1.5$ and (e) different $R$ with $\bar{b} = 1/\kappa^{1/3}$, $g_{\text{th}} = \kappa$.}
    \label{fig_stability_contours}
\end{figure*}

In Fig.~\ref{fig_stability_contours}(a), we show the variation of the stability eigenvalues $\Re (\sigma_j)$ with $\epsilon$ at different values of $\gamma_{||}$. Note that a constant zero eigenvalue emerges for every $\gamma_{||}$ due to the fact that the coupled-mode equations are invariant under a global phase rotation. However, the solution for $\gamma_{||}/\kappa = 100$ is still stable across the entire range of $\epsilon$ since $\Re (\sigma_j) < 0$ for the remaining five eigenvalues. This is in contrast with other systems with optical injection, as in \cite{Zhou2016}, where instability emerges even with infinitely fast inversion relaxation. The solution is marginally stable at the exceptional point ($\epsilon \rightarrow 0$), where a another eigenvalue approaches the imaginary axis. However, the marginal stability disappears once a detuning is applied.

When we decrease $\gamma_{||}/\kappa$ to $1$, the system becomes unstable for $\epsilon/\kappa < 4.3 \times 10^{-2}$, where it has two complex conjugate eigenvalues with real parts greater than zero. The corresponding time-domain fluctuations in lasing frequency at either side of this transition are shown in Fig.~\ref{fig_stability_contours}(b) and (c). This boundary value of $\epsilon$ imposes a minimum detuning below which sensor readout is not feasible. Counterintuitively, this minimum detuning decreases with further decreases in $\gamma_{||}$. For $\gamma_{||}/\kappa = 0.01$, the system is only unstable for $\epsilon/\kappa < 6.3 \times 10^{-3}$.

Figures~\ref{fig_stability_contours}(d) and (e) show the boundary between stable and unstable regions at different signal scale factors $\bar{b}$ and different threshold ratios ($R \equiv g_c/g_{\text{th}}$), respectively. For sufficiently fast relaxation rates, the system is stable regardless of detuning, unless there is a large asymmetry in passive losses (as shown in Fig.~\ref{b_stability_comparison}). This boundary $\gamma_{||}$ value is more strongly influenced by $\bar{b}$, with higher values requiring faster $\gamma_{||}$ to be stabilized. For lower relaxation rates, however, we observe that there is actually a maximum $\gamma_{||}$ to maintain stability that is approximately proportional to the detuning. In other words, continually decreasing the relaxation rate with a constant detuning can cause the system to go from stable to unstable, then back to stable again.

\begin{figure}[ht]
    \centering
    \includegraphics[width=\linewidth]{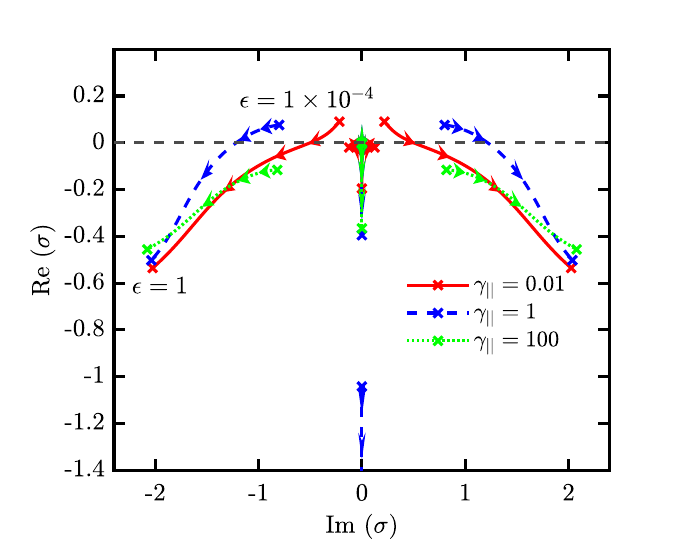}
    \captionsetup{justification=justified, singlelinecheck=false, width=\linewidth}
    \caption{Trajectories of the stability eigenvalues $\sigma_j$ of the coupled resonator system biased at an EP for detuning $\epsilon \in (10^{-4},1)$. The arrows indicate the direction of increasing $\epsilon$. All other parameters are the same as in Fig.~\ref{fig_stability_contours}(a). For slow gain relaxation rate, the system has an unstable complex conjugate eigenvalue pair, which is stabilized as the detuning increases and they pass below the imaginary axis. Since these eigenvalues have the greatest $|\Im (\sigma_j)|$, they describe the highest frequency perturbations applied to the lasing solution.}
    \label{fig_stability_eigs}
\end{figure}

This second transition is due to the fact that the unstable eigenvalues are also those with the highest frequency (largest $|\Im (\sigma)|$). This is shown by looking at the imaginary part of the eigenvalues in Fig.~\ref{fig_stability_eigs}. The oscillation frequency is relevant as the gain medium acts as a low-pass filter. This behavior is intuitively understood by deriving a transfer function from Eq.~\ref{linearized_sa}, which is
\begin{equation}
    F(\omega) = \frac{\delta g_s(\omega)}{\delta (|\psi_1|^2)(\omega)} \propto \frac{\gamma_{||}}{i\omega + \gamma_{||,\mathrm{eff}}},
\end{equation}
where $\gamma_{||,\mathrm{eff}} \equiv \gamma_{||}(1 + |\psi_{1,\text{ss}}|^2)$ is the effective inversion relaxation rate, which depends on the steady-state intensity in the cavity. This is also the corner frequency of the filter.

Increasing the detuning further increases the oscillation frequency of these perturbations, as shown by the eigenvalue trajectories in Fig.~\ref{fig_stability_eigs}, resulting in further damping. This applies whether the complex conjugate eigenvalue pair is stable or unstable.

Based on these results, a coupling rate much slower than the inversion relaxation rate of the gain medium is preferable to maximize stability. However, this limits the responsivity of the system by constraining $\tilde{b}$ beyond the limits imposed by the maximum gain and minimum passive loss of the system (Fig.~\ref{fig_sens}). For slow gain media, it may then be advantageous to choose a coupling rate much larger than $\gamma_{||}$ such that the system remains stable with lower detuning. In either case, designs with $\kappa$ and $\gamma_{||}$ of similar magnitude should be avoided.

\section{SALT Analysis for an Example System}

\begin{figure*}[ht]
    \centering
    
     \begin{minipage}{0.48\textwidth}
        \subfloat{%
            \includegraphics[trim=.2cm .1cm 1.5cm 0cm, clip, width=0.98\linewidth]{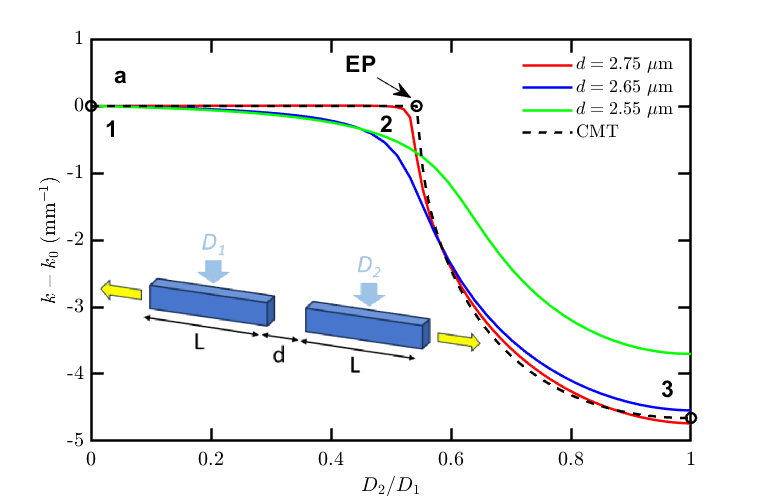}
        }
    \end{minipage}
    \begin{minipage}{0.35\textwidth}
        \subfloat{%
            \includegraphics[trim=.2cm .1cm 1cm 0cm, clip, width=0.98\linewidth]{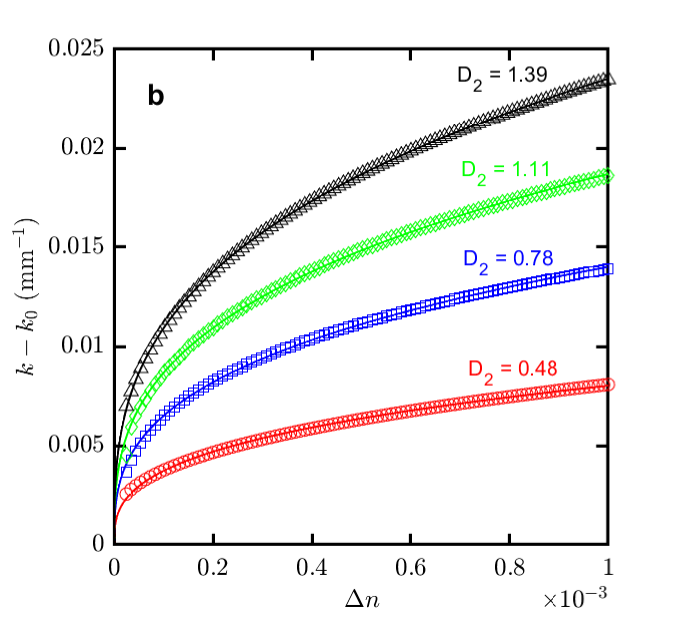}
        }
    \end{minipage}
    \\
    \subfloat{%
        \includegraphics[trim=.0cm 0cm 0cm .05cm, clip, width=0.55\textwidth]{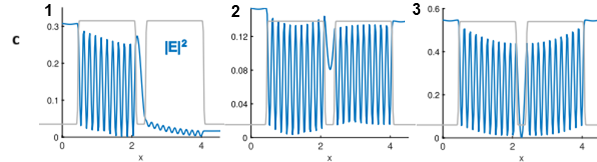}
    }

    \caption{(a) Lasing frequency $k_\mu$ for a single-mode steady-state lasing solution obtained from SALT analysis [Eq.~\ref{salt1}] for different values of gap width $d$. Slab 1 is first pumped up to $D_1 = 0.95$ (equal to 1.2 times its threshold value), then $D_2$ is increased, keeping $D_1$ fixed. A good fit is obtained with the coupled mode theory (CMT) analysis using $\kappa = 1$, $g_0 = 4.75$, and $\gamma_1 = \gamma_2 = 3.55$. The inset illustrates the two-slab configuration with the air gap $d$. (b) Lasing frequency as a function of index perturbation $\Delta n$ applied to one of the slabs for exceptional points located with $D_1$ equal to 1.2, 1.5, 1.8, and 2 times the threshold value. The solid lines are fits with Eq.~\ref{freq_response}. (c) Mode profiles  $|E_\mu(x)|^2$ (blue lines) at different values of $D_2$ showing (1) broken symmetry, (2) near the EP, and (3) unbroken symmetry. The permittivity profiles associated with each mode are also shown in light gray.}
    \label{SALT}
\end{figure*}

In Section~\ref{gainandloss}, we analyzed the frequency response of the system in the presence of gain and/or loss saturation by employing a coupled mode theory (CMT) formalism. While this approach allows us to obtain analytical results, it assumes uniform spatial profiles and neglects the spatially varying gain/loss saturation and the gain-medium line shape.

To probe the validity of our findings in a more general system, we employ Steady-state Ab-initio Laser Theory (SALT) \cite{Esterhazy2014}, which incorporates these spatial effects. For simplicity, we model a laser consisting of two coupled dielectric slabs with non-saturated pumping strengths $D_1$ and $D_2$ (shown in the inset of Fig.~\ref{SALT}(a)). Similar systems have been studied in \cite{ge2016, benzaouia2022}. This configuration corresponds to CMT parameters $g_0 > 0$ and $f_0 < 0$.

Each slab is a one-dimensional dielectric of length $L = 16.67$~\textmu m, made of a two-level gain medium characterized by a refractive index $n_c = 3$, central frequency $k_a / 2\pi = 100$~mm$^{-1}$ (corresponding to a wavelength of 10~\textmu m), and gain linewidth $\gamma_{\perp}/2\pi = 10$~mm$^{-1}$. A conductivity loss $\sigma_c = 0.5 k_a$ is included to represent internal losses within the slabs. The slabs are separated by an air gap of width $d$, allowing for optical coupling between them. 

We track single-mode lasing solutions at frequencies $k_{\mu}$ as the pumping strength is increased by numerically solving the nonlinear steady-state equation

\begin{equation} \label{salt1}
\left[\nabla^2 + k_\mu^2\left( \varepsilon_c(\mathbf{x})+i\frac{\sigma_c}{k_\mu} + \gamma\left(k_\mu\right) D(\mathbf{x}) \right)\right] E_\mu(\mathbf{x})=0,
\end{equation}

where $\varepsilon_c(\mathbf{x})$ represents the spatially varying dielectric constant of the cavity, and $E_\mu(\mathbf{x})$ is the electric field of the $\mu$-th lasing mode. The term $\gamma\left(k_\mu\right) \equiv \gamma_{\perp} / \left(k_\mu - k_a + i \gamma_{\perp}\right)$ is the Lorentzian gain curve of the medium, and $D(\mathbf{x}) = D_0(\mathbf{x}) / \left[1 + |\gamma\left(k_\mu\right) E_\mu(\mathbf{x})|^2\right]$ is the population inversion, which includes the spatial hole-burning effect. Here, $D_0(\mathbf{x})$ represents the non-saturated pumping profile, with $D_0(\mathbf{x}) = D_1$ in slab 1 and $D_0(\mathbf{x}) = D_2$ in slab 2.

Below threshold, where $|E_\mu(\mathbf{x})| \approx 0$, the system can be transformed into a cubic eigenvalue problem by multiplying Eq.~\ref{salt1} by $1/\gamma\left(k_\mu\right)$. The spatial derivatives are discretized using a finite difference scheme. Above threshold, Newton's method is used to track the lasing mode.

Following our CMT analysis in Section~\ref{gainandloss} for $f_0 < 0$, we expect an EP to exist above threshold if $\gamma_2 > \kappa$ and $\gamma_2 < \kappa - f_0$. This can be achieved in the SALT model by appropriately tuning $\sigma_c$ to satisfy these conditions. To reach the EP, we first fix the pump strength $D_1$ (which corresponds to $g_0$ in the CMT model) at a value above its lasing threshold, effectively setting the gain in slab 1. Then, we incrementally increase $D_2$ (which corresponds to $-f_0$), adjusting the gain in slab 2. By varying $D_2$, we control the balance of gain and loss in the system, allowing us to approach the EP. Additionally, we fine-tune the air gap width $d$ to adjust the coupling between the slabs, as the assumption of spatially uniform gain/loss in Eq.~\ref{g_c} is not valid here due to spatial variations in the mode profiles (as evident from the modal intensities in Fig.~\ref{SALT}(c)) that influence the gain through $D(\mathbf{x})$.

At a specific gap width $d = 2.742$~\textmu m, we find a sharp bifurcation in the lasing frequency at a particular value of $D_2$, as expected from Eq.~\ref{vals}. At nearby values of $d$, we observe a similar trend, but the bifurcation becomes smoother. To fine-tune the location of the EP, we maximize the sensitivity $|\frac{dk}{d n}|$ by varying $d$ and $D_2$ in the vicinity of this bifurcation. From this optimized point, a refractive index perturbation $\Delta n$ is applied to the right slab by modifying the permittivity function $\varepsilon_c(\mathbf{x})$. Figure~\ref{SALT}(b) shows that the lasing frequency exhibits a cube-root response to this perturbation, in agreement with our CMT predictions.

\begin{figure*}[!htbp]
    \centering
    \subfloat{%
        \includegraphics[trim=0cm 0cm 1cm .4cm, clip, width=0.35\textwidth]{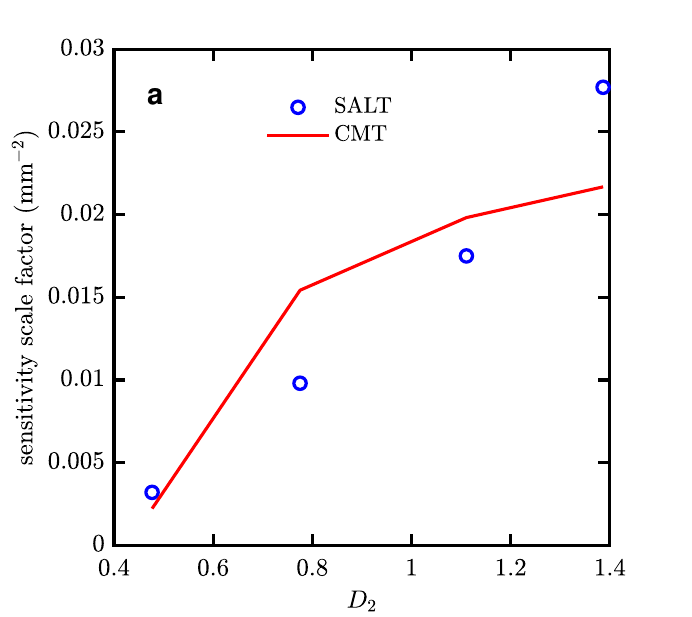}
    }%
    \subfloat{%
        \includegraphics[trim=0cm 0cm 1cm .4cm, clip, width=0.35\textwidth]{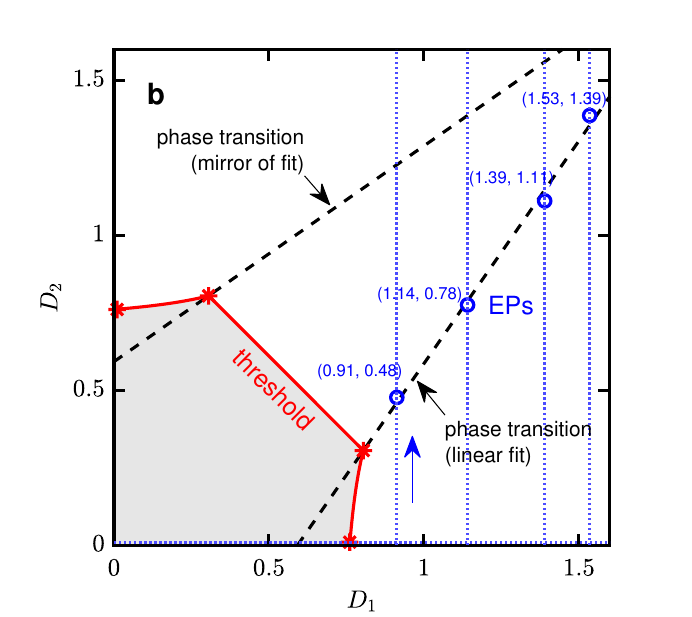}
    }%

     \caption{(a) Signal scale factor (in SALT units) extracted from Fig.~\ref{SALT}(b) compared to best fit from CMT analysis. (b) Threshold map in terms of pumping strengths $D_1$ and $D_2$, constructed based on SALT results. The threshold boundary is inferred from the red asterisks, which are measurements of the lasing threshold. We see that the locations of the EPs fall along a straight line in this space, but that this line does not intersect the origin, as predicted by CMT (Fig.~\ref{fig_threshold_map} (b)).}
    \label{SM_salt}
\end{figure*}

We repeat this procedure at several increasing values of $D_1$, corresponding to different levels above threshold. The corresponding values of $D_2$ where the EP is found are listed in Fig.~\ref{SALT}(b). The scale factor applied to the cube-root response increases with increasing $D_2$, in qualitative agreement with CMT. However, Fig.~\ref{SM_salt}(a) shows that this scaling is approximately linear, whereas CMT predicts sub-linear scaling.

To explain this discrepancy, we create a similar map of the lasing thresholds and phase transitions of the system as in Fig.~\ref{fig_threshold_map}, by assuming that $D_1 \propto g_0$ and $D_2 \propto -f_0$. We find that the locations of the EPs fall along a straight line, but this line does not intersect the origin, contrary to the prediction of Eq.~\ref{g_c}. This could be due to the passive losses of the system being dependent on the mode profiles in the two slabs, leading to different locations for the phase transition above and below threshold.

Another explanation stems from the fact that output coupling loss is not fully accounted for in the CMT analysis. The time reversal of this lasing mode at threshold corresponds to a coherent perfect absorber mode \cite{chong2010} with purely incoming waves from outside the two slabs. Therefore, the lasing mode does not strictly satisfy the PT-symmetry condition $\mathscr{PT} \, E_\mu(\mathbf{x}) = E_\mu^*(-\mathbf{x}) = E_\mu(\mathbf{x})$, even when the modal gain and loss are balanced \cite{ge2016}. This leads to a slightly asymmetric intensity distribution at the locations of the EPs (see profile 2 of Fig.~\ref{SALT}(c) for example). This description applies at and above threshold, but not below threshold where there is zero outgoing radiation. If the output coupling loss is weak compared to the conductivity loss, we expect this symmetry to be recovered \cite{ge2016}. On the other hand, this discrepancy becomes more significant if the system is configured with saturable loss in the second slab, since we expect that the conductivity loss would have to be decreased such that the total passive losses do not exceed the coupling rate. In future work, this regime could instead be probed by using distributed Bragg reflectors (DBRs) to reduce the output coupling loss.

\section{Discussion}

In summary, we have shown, using coupled mode theory, that the steady-state output frequency of a laser biased near the PT-symmetric EP is modified proportionally to the cube root of the perturbation strength. Using SALT analysis, we have shown that this response can still exist in a system where the mode/gain profiles are inhomogeneous, but the tuning process is made more demanding. The constant of proportionality—a signal scale factor—is determined by the relative strength of the saturable gains and/or losses in each resonator. Unlike previous proposals that utilized only one nonlinear resonator \cite{Bai2022}, our approach leverages the interplay between two saturable resonators to achieve an increase in responsivity of several orders of magnitude. We also show that maximizing this scale factor makes the system more robust to parametric errors. Such errors in gain and/or coupling rate make it challenging to bias the system exactly at the EP, even when both resonators are in tune.

At first glance, the form of this response appears to contradict previous experimental results using a seemingly comparable system \cite{Hodaei2016}, which exhibited a square-root response. This is explained by the difference in measurement schemes: if the pump pulse duration is short compared to the gain relaxation time, the square-root response is readily observed. However, the cube-root response emerges when the duration is extended such that the steady-state solution is reached. This is shown by looking at the time evolution of the system under different gain relaxation rates in Supplementary Section~III.

In this work, we have only considered parametric errors which vary much more slowly than the system dynamics or the measurement rate. Therefore, even though we have shown an enhancement in the lasing frequency shift (responsivity) at a nonlinear EP, we cannot claim that this comes with an improvement in sensing precision (sensitivity), since the latter is affected by noise and linewidth considerations. A full discussion of parametric and quantum noise in the vicinity of a nonlinear EP is left for future work. Several works \cite{Chen2019, Wang2020, Mortensen2018, Zhang2018} cast doubt on the idea that precision can be improved by sensing at linear exceptional points due to Petermann linewidth broadening that compensates for the increase in SEF. However, studies of fundamental noise and precision limits in nonlinear systems are still limited \cite{Silver2021, Bai2022}, potentially leaving the door open for enhanced precision to be demonstrated under certain circumstances.

We also defined the regions where the EP state becomes unstable, leading to either limit cycles or evolution into a different stable state with broken symmetry. When accounting for gain dynamics and asymmetry in both linear and nonlinear gain/loss, there are still wide parameter regions where lasing at a stable EP is possible. By identifying these stable parameter regions and demonstrating enhanced responsivity, our work lays the groundwork for developing highly responsive sensors that operate under realistic conditions, despite inherent system imperfections.

\FloatBarrier

\appendix

\section{Derivation and Stability Analysis of Lasing Modes at Zero Detuning}
\label{non_degenerate_modes}

The saturated gain and loss, $g_s$ and $f_s$, for degenerate lasing modes are restricted to a single value, corresponding to the PT-symmetric EP, which emerges at $g_0 = g_c$.

On the other hand, $g_s$ and $f_s$ for non-degenerate modes can have multiple solutions.

We construct a system of four equations for $g_s$, $f_s$, $I_1$, and $I_2$ by using Eqs.~\ref{g_s} and \ref{f_s} along with

\begin{equation}\label{cond3}
    (g_s - \gamma_1)(f_s + \gamma_2) = \kappa^2,
\end{equation}

and

\begin{equation}\label{cond4}
    \frac{I_1}{I_2} = \left(\frac{\kappa}{g_s - \gamma_1}\right)^2,
\end{equation}

where Eq.~\ref{cond3} comes from the condition that $\Im (\omega - \omega_0) = 0$, and Eq.~\ref{cond4} is derived from the eigenvector equation (Eq.~\ref{vects}).

These conditions yield a quartic equation for $g_s$:

\begin{equation} \label{quartic}
\begin{aligned}
    P(g')
    &= \gamma_2 (g')^4 + (\gamma_1 \gamma_2 - g_0 \gamma_2 - \kappa^2) (g')^3 \\
    &\quad + \kappa^2 (g_0 - \gamma_1 - \gamma_2 - f_0) (g')^2 \\
    &\quad + \kappa^2 \left(\kappa^2 - \gamma_1 (f_0 + \gamma_2)\right) g' + \gamma_1 \kappa^4 = 0.
\end{aligned}
\end{equation}

The PT-symmetric EP $g' = f' = \kappa$ is always a solution when $g_0 = g_c$ (Eq.~\ref{g_c}), provided that $f_0$ and $\gamma_2$ are chosen such that the EP exists at or above threshold. Depending on the exact values of the parameters, $P(g')$ can have either one or three other real roots. However, only real roots corresponding to positive $I_1$ and $I_2$ can represent physical solutions.

In previous work studying the system with symmetric passive losses \cite{Hodaei2016, benzaouia2022}, for instance, these additional solutions were neglected. It is straightforward to show that if an EP exists above threshold—either with $f_0 > 0$ or $f_0 < 0$—and $\gamma_1 = \gamma_2$, then $P(g')$ remains positive and monotonically increasing for $g' \in (0, \kappa)$. This eliminates the possibility of an additional real root that would be energetically favored over the PT-symmetric EP.

However, this is not generally the case with asymmetric losses: if an EP can exist above threshold with $f_0 > 0$, there can be a real root $g' \in (0, \kappa)$ if $\gamma_1$ is substantially greater than $\gamma_2$. Similarly, with $f_0 < 0$, there is a narrow range of parameters where a root can exist if $\gamma_2$ is substantially greater than $\gamma_1$. An example of the roots of the system, as a function of this loss difference, is shown in Fig.~S1.

We conducted a linear stability analysis, which confirmed that the EP solution is unstable if another physical solution with $g' \in (0, \kappa)$ exists (not shown). We used the same approach as in Appendix~\ref{linearized_sa}, but focused on the limit of fast inversion relaxation rate ($\gamma_{||} = 10^6 \kappa$).

\section{Derivation of the Signal Scale Factor} \label{sens_derivation}

To explicitly solve for the signal scale factor $\tilde{b}$ for the general case where $f_0 \neq 0$, we aim to ensure consistency between the intensity ratios described by the eigenvector equations and the saturation equations.

By substituting Eqs.~\ref{g_u} and \ref{f_u} into Eqs.~\ref{g_s} and \ref{f_s}, we obtain:

\begin{equation} \label{R_sat}
    \left| \frac{\psi_1}{\psi_2} \right|^2 \approx 1 + \frac{f_0 \left[ (\bar{b} - \delta) \gamma_1 + (\bar{b} + \delta) \gamma_2 - 2 \delta \kappa \right]}{(\kappa + \gamma_1)(\kappa - \gamma_2)(\kappa - f_0 - \gamma_2)} \epsilon^{2/3}.
\end{equation}

The solution based on these saturation equations must be consistent with the system eigenvectors. Keeping only lowest-order terms, Eq.~\ref{vects} yields:

\begin{equation} \label{R_vec}
    \left| \frac{\psi_1}{\psi_2} \right|^2 \approx 1 + \frac{2 \delta}{\kappa} \epsilon^{2/3}.
\end{equation}

Substituting Eq.~\ref{b} into Eq.~\ref{R_sat} and equating it with Eq.~\ref{R_vec} yields a cubic equation for $\delta$. The real solution is

\begin{equation}
\delta = \frac{\kappa}{2} \left( \frac{f_0 (\gamma_1 + \gamma_2)}{\kappa T} \right)^{1/3},
\end{equation}

where

\begin{equation} \label{eq:T_simplified_appx}
T = 2 (\kappa + \gamma_1)(\kappa - \gamma_2)^2 - f_0 \left[ \kappa (\gamma_1 - \gamma_2) - 2 \gamma_1 \gamma_2 \right ].
\end{equation}

The signal scale factor $\tilde{b} = \kappa \bar{b}$ can then be found using Eq.~\ref{b}.

\section{Relation Between Signal Scale Factor and Stability} \label{b_stability}

In Section~\ref{gainandloss}, we found that the signal scale factor can take on zero or complex values for certain parameter sets, including some parameter sets that yield a physical EP above threshold. In this section, we will show that (with some exceptions) the conditions for real $\tilde{b}$ are equivalent to the conditions for stability at the PT-symmetric EP, which are outlined in Appendix~\ref{non_degenerate_modes}.

We begin by looking at $P(g')$ (Eq.~\ref{quartic}) biased at $g_0 = g_c$. Using Eq.~\ref{eq:T_simplified}, the $f_0$ terms can be replaced by

\begin{equation} \label{f_0_T}
f_0 = \frac{T - (\gamma_2 - \kappa)^2 (2\gamma_1 + 2\kappa)}{\kappa (\gamma_2 - \gamma_1) + 2\gamma_1 \gamma_2},
\end{equation}

which yields $P(g') = \frac{(g' - \kappa) N}{(\gamma_2 - \kappa) D}$, where

\begin{align}
N &= C T + (g' - \kappa)(\gamma_2 - \kappa)(a {g'}^2 + b g' + c), \\
D &= \kappa(\gamma_2 - \gamma_1) + 2\gamma_1 \gamma_2,
\end{align}

with

\begin{align*}
C &= \gamma_2 {g'}^2 \kappa + \gamma_1 \gamma_2 {g'}^2 - \gamma_1 g' \kappa^2 + \gamma_1 \gamma_2 g' \kappa, \\
a &= 2 \gamma_1 \gamma_2^2 - \gamma_1 \gamma_2 \kappa + \gamma_2^2 \kappa, \\
b &= \gamma_1^2 \gamma_2 \kappa + \gamma_1 \gamma_2^2 \kappa + \gamma_1 \kappa^3 + \gamma_2 \kappa^3, \\
c &= 2 \gamma_1^2 \gamma_2 \kappa^2 - \gamma_1^2 \kappa^3 + \gamma_1 \gamma_2 \kappa^3.
\end{align*}

The one root of $P$ at $g' = \kappa$ corresponds to the PT-symmetric EP. We are interested in the stability at $\tilde{b} = 0$ (requiring $T = 0$), since this separates regions where the scale factor is real from those where it is complex. We will therefore proceed by analyzing the roots of Eq.~\ref{quartic} at $T = 0$. We can then analyze the movement of roots under the effect of a small, non-zero $T$.

At $T = 0$, a second root of $P$ at $g' = \kappa$ emerges. This intersection is shown numerically in Figure~S1. Also at this point, we get

\begin{equation}
\left.\frac{d g'}{d T}\right|_{T=0,\, g'=\kappa} = -
C \left.\left(\frac{\partial N(g', T)}{\partial g'}\right)^{-1} \right|_{T=0,\, g'=\kappa},
\end{equation}

where

\begin{multline}
\left.\frac{\partial N(g', T)}{\partial g'}\right|_{T=0,\, g'=\kappa} = \kappa^2 (\gamma_1 + \gamma_2) (\gamma_2 - \kappa) \\
\times \left(3 \gamma_1 \gamma_2 - \gamma_1 \kappa + \gamma_2 \kappa + \kappa^2\right).
\end{multline}

Therefore, the new location for this root of $N$, $P$ is

\begin{equation}
g'(T) \approx \kappa + T \left.\frac{d g'}{d T}\right|_{T=0,\, g'=\kappa}.
\end{equation}

For $f_0 < 0$, $\gamma_2 > \kappa$, we get $C > 0$ and $\frac{\partial N(g', T)}{\partial g'} > 0$ such that $\frac{d g'}{d T}$ is negative. Therefore, introducing a small $T > 0$ shifts the location of the root to a lower value such that $0 < g'(T) < \kappa$. This results in the EP mode becoming unstable due to gain competition. Since $f_0 < 0$, this also results in the radicand of Eq.~\ref{b} being negative and $\tilde{b}$ being complex. Conversely, $T < 0$ results in real $\tilde{b}$ since this root is shifted to higher modal gain.

The conditions for the signal scale factor to be real with $f_0 < 0$ are therefore equivalent to the conditions for the EP mode to be stable.

On the other hand, if $f_0 > 0$, $\gamma_2 < \kappa$, we get that $C$ is usually positive and $\frac{\partial N(g', T)}{\partial g'}$ is usually negative. These signs can both change in the case where $\gamma_1$ is much larger than $\gamma_2$ and $\kappa$. In either case, we get that $\frac{d g'}{d T}$ is positive. Therefore, $T > 0$ shifts the root to higher modal gain while ensuring that $\tilde{b}$ is real. However, in this case, real $\tilde{b}$ does not guarantee the stability of the EP mode.

For completeness, we must consider the movement of the other two roots to ensure that $P$ has no other real roots on the interval $(0, \kappa)$. We have already described the behavior of two roots: the root of $P$ that remains at $g' = \kappa$ and the root of $N$ which is at $\kappa$ when $T = 0$, but does not remain there for $T \neq 0$. The remaining roots of $P$ are the two roots corresponding to the roots of $a {g'}^2 + b g' + c$ when $T = 0$.

For $f_0 < 0$, we get that $a > 0$ and $b > 0$, so the quadratic opens upwards. Its minimum value on the interval $(0, \kappa)$ is therefore either at the vertex or at the boundary of the interval. However, the vertex at $g' = -b/2a$ is outside of the interval. The minimum value at $g' \rightarrow 0^{+}$ is $c$, which is also positive. We can therefore conclude that both roots are negative. For most parameter values, we find that the location of this vertex is sufficiently negative that these roots remain negative over a wide range of $T$.

However, for $f_0 > 0$, we get that $a$ is usually positive but can be negative for large values of $\gamma_1$. Meanwhile, $b$ is always positive. Therefore, it is possible for the vertex—or one of the roots—to be in $(0, \kappa)$. An example is provided in Figure~S1.

\section{Linear Stability Analysis} \label{linear_stability_analysis}

We combine Eqs.~\ref{gaindynamics} and~\ref{lossdynamics} with Eq.~\ref{shrodinger}, and linearize the system for small perturbations about the steady-state lasing modes by inserting

\begin{subequations}
    \begin{align}
    \psi_1(t) &= [\psi_{1,\text{ss}} + \delta \psi_1(t)] e^{-i \omega t}, \\
    \psi_2(t) &= [\psi_{2,\text{ss}} + \delta \psi_2(t)] e^{-i \omega t}, \\
    g_s(t) &= g_{s,\text{ss}} + \delta g_s(t), \\
    f_s(t) &= f_{s,\text{ss}} + \delta f_s(t),
    \end{align}
\end{subequations}

where $e^{-i \omega t} \left( \psi_{1,\text{ss}}, \psi_{2,\text{ss}} \right)^\text{T}$ is the solution to Eq.~\ref{shrodinger} corresponding to the real eigenvalue $\omega$, expressed in the rotating frame at $\omega_0$.

We collect terms by order in the perturbation $\delta$. The first-order equations are

\begin{subequations} \label{linearized_sa}
    \begin{align}
        \frac{d (\delta \psi_1)}{dt} &= \left( g' + i(\omega - \epsilon) \right) \delta \psi_1 + \kappa \delta \psi_2 + \psi_{1,\text{ss}} \delta g_s, \\
        \frac{d (\delta \psi_2)}{dt} &= \left( -f' + i \omega \right) \delta \psi_2 + \kappa \delta \psi_1 - \psi_{2,\text{ss}} \delta f_s, \\ 
        \frac{d (\delta g_s)}{dt} &= - \gamma_{||}(1 + |\psi_{1,\text{ss}}|^2) - 2 g_s \gamma_{||} \delta (|\psi_1|^2), \\
        \frac{d (\delta f_s)}{dt} &= - \gamma_{||}(1 + |\psi_{2,\text{ss}}|^2) - 2 f_s \gamma_{||} \delta (|\psi_2|^2),
    \end{align}
\end{subequations}

where

\begin{equation}
    \delta (|\psi_i|^2) = \Re (\psi_{i,\text{ss}}) \Re (\delta \psi_i) + \Im (\psi_{i,\text{ss}}) \Im (\delta \psi_i),
\end{equation}

for $i = 1, 2$. Splitting Eqs.~\ref{linearized_sa}~(a) and~(b) into their real and imaginary parts gives rise to a set of linear equations for six variables. For convenience, we can combine them into a single unknown vector:

\begin{equation}
\mathbf{u}(t) = \begin{pmatrix}
\Re \delta \psi_1(t) \\
\Im \delta \psi_1(t) \\
\Re \delta \psi_2(t) \\
\Im \delta \psi_2(t) \\
\delta g_s(t) \\
\delta f_s(t)
\end{pmatrix} = \Re \left( \mathbf{x} e^{\sigma t} \right),
\end{equation}

so that Eq.~\ref{linearized_sa} can be written as a standard eigenvalue problem using the Jacobian $A$:

\begin{equation}
    (A - \sigma I) \mathbf{x} = 0.
\end{equation}

This eigenvalue problem can be solved numerically, resulting in a set of eigenvalues and eigenvectors $\{ \sigma_j, \mathbf{x}_j \}$. Eigenvalues $\Re(\sigma_j) > 0$ describe perturbations that grow exponentially in time, implying that the system is unstable. Due to the invariance of our system to global phase rotation, we permit one eigenvalue $\Re(\sigma_i) = 0$. If $\Re(\sigma_j) < 0$ for every other eigenvalue $j \neq i$, the solution is stable against small perturbations.

\bibliographystyle{apsrev4-2}
\bibliography{mybib_abbrev_noeprint_nov1_v2}%

\end{document}


\preprint{APS/123-QED}

\title{Supplementary Material: Responsivity and Stability of Nonlinear Exceptional Point Lasers with Saturable Gain and Loss}

\author{Todd Darcie}
\affiliation{Department of Electrical and Computer Engineering\\University of Toronto.}
\email{todd.darcie@mail.utoronto.ca}

\author{J. Stewart Aitchison}
\affiliation{Department of Electrical and Computer Engineering\\University of Toronto.}

\date{\today}

\maketitle

\section{Mode stability with asymmetric losses}

\begin{figure*}[!htbp]
    \centering
    \subfloat{%
        \includegraphics[trim=0cm 0cm 1cm .4cm, clip, width=0.35\textwidth]{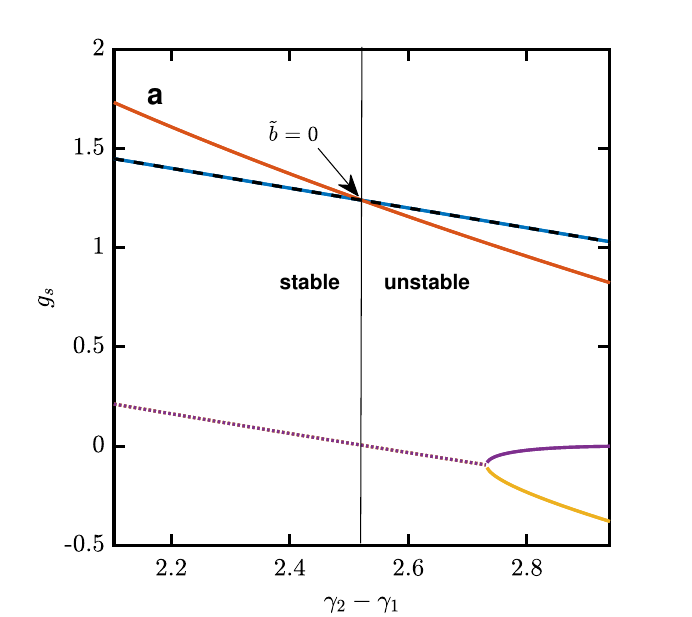}
    }%
    \subfloat{%
        \includegraphics[trim=0cm 0cm 1cm .4cm, clip, width=0.35\textwidth]{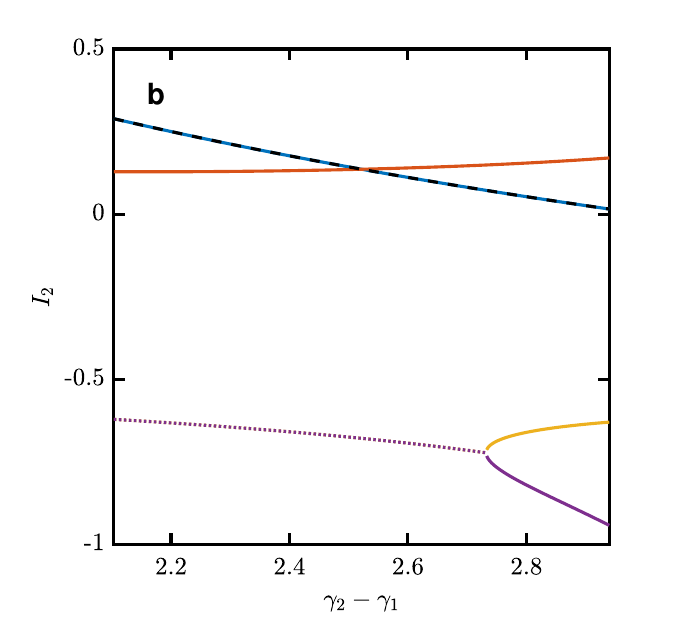}
    }%
    \\
    \subfloat{%
        \includegraphics[trim=0cm 0cm 1cm .4cm, clip, width=0.35\textwidth]{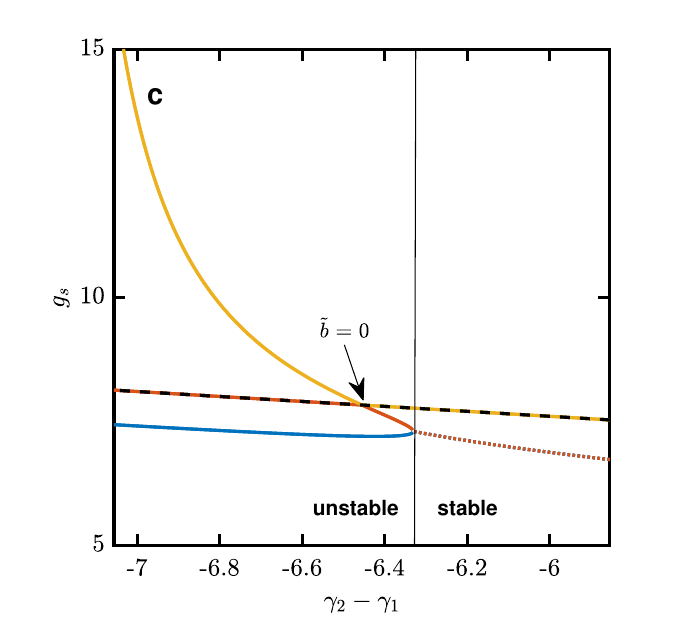}
    }%
    \subfloat{%
        \includegraphics[trim=0cm 0cm 1cm .4cm, clip, width=0.35\textwidth]{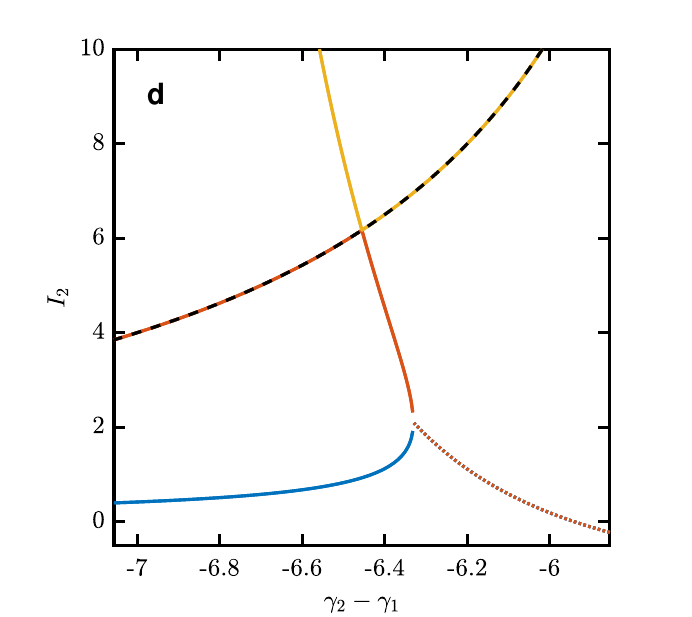}
    }%

     \caption{Solutions to Eq.~A3 as a function of the asymmetry in passive losses $\gamma_2 - \gamma_1$: (a) Saturable gain and (b) intensity in resonator 2 for $f_0 = -2$, $\gamma_{\text{avg}} \equiv (\gamma_1 + \gamma_2)/2 = 1.5$, and $\kappa = 1$. The solution corresponding to the PT-symmetric EP is shown with the dashed black line. Regions where the parameters are complex are shown with dotted lines. We see that the EP solution has the lowest gain of all physical solutions (with intensities real and non-negative) until the asymmetry passes a critical value, which corresponds to the point where the sensitivity scale factor $\tilde{b} = 0$. This condition is required to maintain stability at the EP. (c) Saturable gain and (d) intensity in resonator 2 for $f_0 = 4.5$, $\gamma_{\text{avg}} = 3.6$, and $\kappa = 1$. In this case, the critical asymmetry has the opposite sign. Also, the point where $\tilde{b} = 0$ no longer corresponds with this stability boundary, which instead coincides with the bifurcation where the system goes from four to two real roots (intersection of red and blue lines).}
    \label{SM_mode_stability}
\end{figure*}

\FloatBarrier

\section{\label{sec:param_errors} Impact of Parametric Errors}

We first define $r$ to describe the error between the physical coupling rate and the coupling rate corresponding to the EP: 

\begin{equation}
    r \equiv \frac{\kappa - \kappa_{\mathrm{EP}}}{\kappa_{\mathrm{EP}}}.
\end{equation}

In terms of $g_0$, this error is 

\begin{equation}
    r = \frac{f_0 (1 + \gamma_1) - g_0 (1 - \gamma_2)}{g_0 - f_0},
\end{equation}
such that $r = 0$ if $g_0 = g_c$. For convenience, we have also normalized rates $f_0, g_0, \gamma_1, \gamma_2, \omega_0$ and $\epsilon$ with respect to $\kappa_{EP}$ for the rest of this section. The eigenvalues of the system are 

\begin{equation} \label{lambda_sm}
    \omega_{1,2} =  \omega_0 + \frac{i(g' - f')}{2} + \frac{\epsilon}{2} \pm \frac{i \sqrt{x - i y}}{2},
\end{equation}

where $x = (g' + f')^2 - 4 (1 + r)^2 - \epsilon^2$ and $y = 2 \epsilon (g' + f')$. 

We consider saturable gain/loss of the form 

\begin{equation} \label{g_prime_sup}
    g' = g'(0) - \beta_g |\epsilon|^s,
\end{equation} 

\begin{equation}\label{f_prime_sup}
    f' = f'(0) - \beta_f |\epsilon|^s,
\end{equation} 

where we assume that $s$ is positive. We are interested in finding the lowest power of $\epsilon$. 

The real and imaginary parts of Eq.~\ref{lambda_sm} can be found with the identities 

\begin{equation}\label{real_ident}
\Re\left(\sqrt{x - i y}\right) = \left(x^2 + y^2\right)^{1/4} \cos\left(\frac{1}{2} \arg(x - i y)\right),
\end{equation}

\begin{equation}\label{imag_ident}
\Im\left(\sqrt{x - i y}\right) = \left(x^2 + y^2\right)^{1/4} \sin\left(\frac{1}{2} \arg(x - i y)\right).
\end{equation}

We identify two regimes based on whether $x$ is greater than or less than zero. Since we are looking for the maximum SEF, which is reached as the detuning approaches zero, we assume $|r| \gg |\epsilon|$ such that $x < 0$ when $r > 0$ and vice versa. 

\subsection{Unbroken Symmetry ($r > 0$)}

First, we look at $\epsilon = 0$. Equation~\ref{lambda_sm} gives

\begin{equation} \label{im_unbroken}
    2 \Im\left(\omega_{1,2}-\omega_0\right) = g'(0) - f'(0),
\end{equation}

\begin{equation}
    2 \Re\left(\omega_{1,2}-\omega_0\right) = \pm \left|\Gamma'\right|^{1/2},
\end{equation}

where $\Gamma' = \left[ \left( g'(0) + f'(0) \right)^2 - 4 (1 + r)^2 \right]$. Since the imaginary parts of the eigenvalues are degenerate, this corresponds to unbroken symmetry. Accordingly, the eigenvector equation for the lasing mode (main text Eq.~7) gives

\begin{equation} \label{vect_unbroken}
       \left| \frac{\psi_1}{\psi_2} \right|^2 \approx \frac{4 f'(0)^2 + |\Gamma'|}{4 (1 + r)^2} =  1.
\end{equation}

Together, Eqs.~\ref{im_unbroken} and \ref{vect_unbroken} give $g'(0) = f'(0) = 1$. 
We now consider $\epsilon \neq 0$. Keeping only lowest-order terms in $\epsilon$ gives 

\begin{equation}
    \frac{y}{x} \approx \frac{4 \epsilon}{|\Gamma'|},
\end{equation}

\begin{equation}
    x^2 + y^2 \approx (\Gamma')^2,
\end{equation}

where the same terms are kept so long as $s > 0$. 

According to Eqs.~\ref{lambda_sm}, \ref{real_ident}, and \ref{imag_ident}, the real and imaginary parts of the system eigenvalues for $x < 0$ in the limit $y/x \ll 1$ are 

\begin{equation}
    2 \Re\left( \omega - \omega_0 \right) \approx \epsilon \pm \left[ |\Gamma'|^{1/2} + \frac{2 |\epsilon| (1 - |\Gamma'|) (\beta_g + \beta_f)}{|\Gamma'|^{1/2}} \right],
\end{equation}

\begin{equation}\label{imag_sup}
    2 \Im\left( \omega - \omega_0 \right) \approx (g' - f') \pm \frac{2 |\epsilon|}{|\Gamma'|^{1/2}}.
\end{equation} 

For one of the eigenvalues to be entirely real requires that the gain and loss saturate proportionally to $|\epsilon|$, such that $s = 1$. Similarly, equating linear terms requires that

\begin{equation}
    \delta_{\beta} \equiv \frac{\beta_f - \beta_g}{2} = |\Gamma'|^{-1/2}.
\end{equation}

If $r, |\Gamma'| \ll 1$, the first and last terms can be discarded, so the $SEF$ becomes
\begin{equation} \label{SEF_unbroken}
    SEF \equiv \left| \frac{\partial \Re\left( \omega - \omega_0 \right) }{ \partial \epsilon } \right|^2 \approx \frac{ (\beta_g + \beta_f)^2 }{ |\Gamma'| }.
\end{equation}

For the case where resonator 2 is passive ($f_0 = 0$), we must have $\beta_f = 0$, therefore $\beta_g \approx |\Gamma'|^{-1/2}$. Lastly, $|\Gamma'| \approx 8 r$ for $r > 0$ and $r \ll 1$. Therefore, Eq.~\ref{SEF_unbroken} is 

\begin{equation}
    SEF \approx \frac{1}{|\Gamma'|^2} \approx \frac{1}{64 r^2}.
\end{equation}

For the more general case where $f_0 \neq 0$, the value of $\bar{\beta} \equiv (\beta_g + \beta_f)/2$ can be found by equating the intensity ratio prescribed by the eigenvector of the lasing mode (Eq.~7) and equating it to the intensity ratio prescribed by the saturation equations:

\begin{equation} \label{eig_unbroken}
    \left| \frac{\psi_1}{\psi_2} \right|^2 \approx 1 + 2 (\bar{\beta} + \delta_{\beta} ) |\epsilon|,
\end{equation}

\begin{equation} \label{R_sat}
    \left| \frac{\psi_1}{\psi_2} \right|^2 \approx 1 - \frac{ f_0 \left[ (\bar{\beta} - \delta_{\beta} ) \gamma_1  + (\bar{\beta} + \delta_{\beta} ) \gamma_2 - 2 \delta_{\beta} \right] |\epsilon| }{ (1 + \gamma_1)(1 - \gamma_2)( f_0 + \gamma_2 - 1) }.
\end{equation}

It is straightforward to show that equating $|\epsilon|$ terms requires that $\bar{\beta} = k |\Gamma'|^{-1/2}$, where $k$ is a constant. With some manipulation, we can also show that $k \propto \bar{b}^3$. 

\begin{equation}
    SEF \approx \frac{k}{|\Gamma'|^2}.
\end{equation}

This is shown numerically in the main text. 

\subsection{Broken Symmetry ($r < 0$)}

First, looking at $\epsilon = 0$, Eq.~\ref{lambda_sm} gives 

\begin{equation} \label{im_broken}
    2 \Im\left( \omega_{1,2} - \omega_0 \right) = g'(0) - f'(0) \pm \Gamma'^{1/2},
\end{equation}

\begin{equation}
    2 \Re\left( \omega_{1,2} - \omega_0 \right) = 0,
\end{equation}

where the imaginary parts are non-degenerate, indicating broken symmetry. Accordingly, the eigenvector equation for the lasing mode (main text Eq.~7) gives 

\begin{equation} \label{vect_broken}
       \left| \frac{\psi_1}{\psi_2} \right|^2 \approx \frac{ f'(0)^2 }{ (1 + r)^2 }.
\end{equation}

For $\epsilon \neq 0$, the lowest-order term in $y/x$ is 

\begin{equation}
    \frac{y}{x} \approx \frac{2 (g'(0) + f'(0)) \epsilon}{\Gamma'}.
\end{equation}

However, because $x > 0$, the argument of $x - i y$ is no longer shifted by $\pi$, so the cosine term in Eq.~\ref{real_ident} is

\begin{equation} \label{cos_broken}
    1 - \frac{ ( g'(0) + f'(0) )^2 \epsilon^2 }{ 2 (\Gamma')^2 } + \mathcal{O}(\epsilon^4).
\end{equation}

We also look at the next terms in $x^2 + y^2$, which are 

\begin{multline} \label{quad_broken}
    x^2 + y^2 \approx (\Gamma')^2 + 2 |\epsilon|^2 \left[ 2 (g'(0) + f'(0))^2 - \Gamma' \right] \\
    - 4 \Gamma' |\epsilon|^{s} (\beta_g + \beta_f) (g'(0) + f'(0)) \\
    + |\epsilon|^{2s} (\beta_g + \beta_f)^2 \left[ 4 (g'(0) + f'(0))^2 + 2 \Gamma' \right]
    + \ldots
\end{multline}  

We note that if $s > 2$, such that the second term dominates, then $(x^2 + y^2)^{1/4}$ will be equal to $\Gamma'^{1/2} + \mathcal{O}(|\epsilon|^{2})$, while $g' - f'$ is only of order $|\epsilon|^s$. Since $\beta_g$, $\beta_f$, $g'(0)$, and $f'(0)$ are greater than zero, this would lead to mismatched powers of $\epsilon$. Since the power series for cosine (Eq.~\ref{cos_broken}) contains only even powers, we must then have $s = 0$ or $2$ to avoid introducing unmatched powers in the imaginary part of the eigenvalue. Choosing $s = 0$ results in an imaginary part that is inconsistent with Eq.~\ref{im_broken} for $\epsilon = 0$.

Therefore, we must have $s = 2$ for broken symmetry. With this exponent, applying Eq.~\ref{real_ident} and keeping up to quadratic terms yields

\begin{equation}\label{imag_broken}
    2 \Im\left( \omega - \omega_0 \right) \approx (g' - f') \pm \left( \Gamma'^{1/2} + C |\epsilon|^2 \Gamma'^{-3/2} \right),
\end{equation}

where 
\begin{multline}
    C = -2 \bar{\beta} ( g'(0) + f'(0) )^3 + 8 \bar{\beta} ( g'(0) + f'(0) )(1 + r)^2 \\
    + 2 (1 + r)^2,
\end{multline}

and $\bar{\beta} \equiv (\beta_f + \beta_g)/2$. The real part of the eigenvalue is 

\begin{equation}
    2 \Re\left( \omega - \omega_0 \right) \approx \frac{\epsilon}{2} \left( 1 +  \frac{2 ( g'(0) + f'(0) ) }{ \Gamma'^{1/2} } \right),
\end{equation}

which leads to

\begin{equation}
    SEF \approx \frac{ ( g'(0) + f'(0) )^2 }{ \Gamma' }.
\end{equation}

The relation between $r$ and $\Gamma'$ is determined by equating the intensity ratio prescribed by the eigenvector equations (Eq.~\ref{vect_broken}) and the saturation equations (Eq.~7). Unlike the unbroken regime where $|\Gamma'| \approx 8 r$, the linear term in the broken regime cancels out due to the dependence of $g'(0)$ and $f'(0)$ on $r$, leaving

\begin{equation}
\Gamma' = (A^2 + A + 4) r^2,
\end{equation}

where 
\begin{equation}
A = \frac{16 ( \gamma_2 - 1 ) }{ ( 2 f_0 + \gamma_1 ) ( 1 + \gamma_2 ) }.
\end{equation}

Therefore, $SEF \propto 1 / r^2$, as in the unbroken case. 

\section{Transient Behavior of a Nonlinear Binary System}

\begin{figure*}[!htbp]
    \centering
    \subfloat[]{%
        \begin{overpic}[trim=.5cm 0cm .9cm .5cm, clip, width=0.45\textwidth]{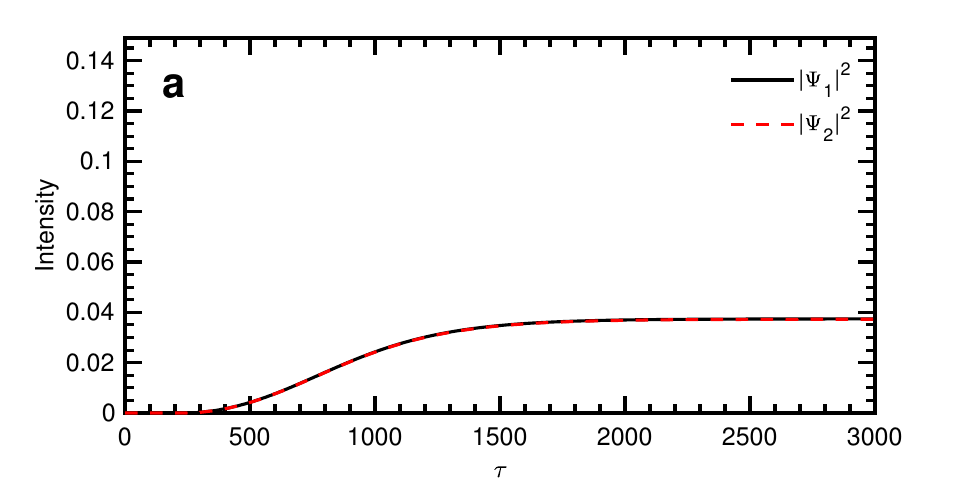}
            \put(50, 0){
                \colorbox{white}{
                    \makebox[0pt][c]{ $\tau$}
                }
            } 
            \put(-2.5, 20){
                \colorbox{white}{
                    \makebox[0pt][r]{\rotatebox{90}{\scriptsize Intensity}}
                }
            }
        \end{overpic}
    }
    \hspace{0.05\textwidth} %
    \subfloat[]{%
        \begin{overpic}[trim=.5cm 0cm .9cm .5cm, clip, width=0.45\textwidth]{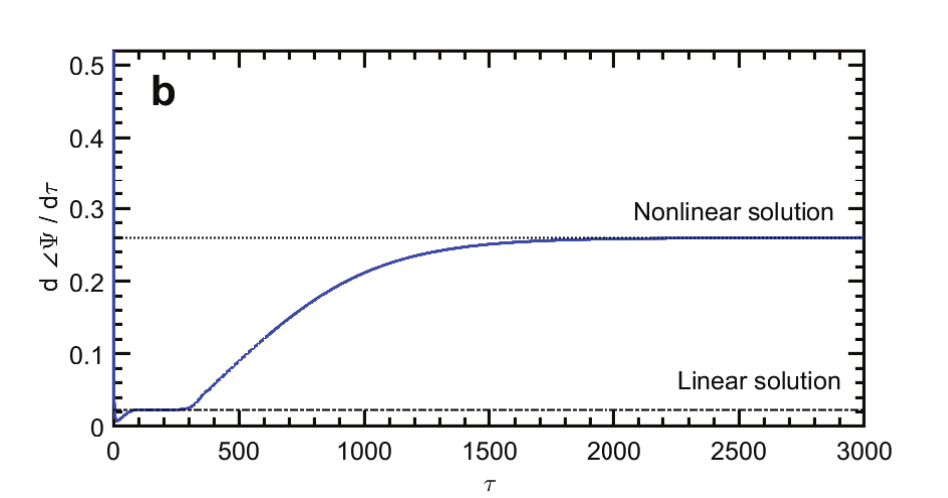}
            \put(50, 0){
                \colorbox{white}{
                    \makebox[0pt][c]{ $\tau$}
                }
            } 
            \put(-1.9, 20){
                \colorbox{white}{
                    \makebox[0pt][r]{\rotatebox{90}{\scriptsize Frequency}}
                }
            }
        \end{overpic}
    }

    \vspace{0.5cm} %

    \subfloat[]{%
        \begin{overpic}[trim=.5cm 0cm .9cm .5cm, clip, width=0.45\textwidth]{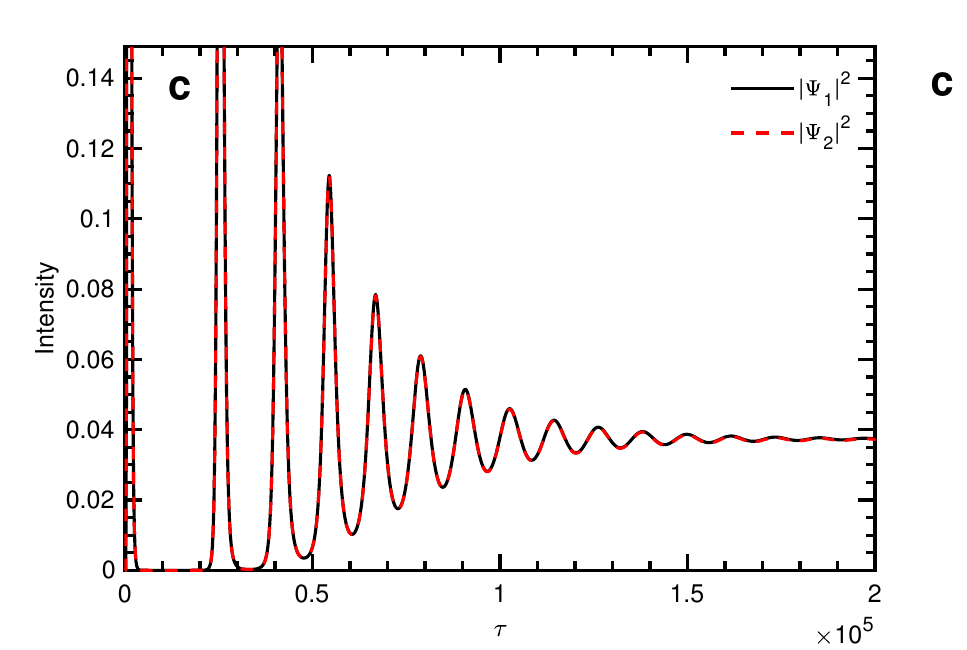}
            \put(50, 0){
                \colorbox{white}{
                    \makebox[0pt][c]{ $\tau$}
                }
            } 
            \put(-2.5, 30){
                \colorbox{white}{
                    \makebox[0pt][r]{\rotatebox{90}{\scriptsize Intensity}}
                }
            }
        \end{overpic}
    }
    \hspace{0.05\textwidth} %
    \subfloat[]{%
        \begin{overpic}[trim=.5cm 0cm .9cm .5cm, clip, width=0.45\textwidth]{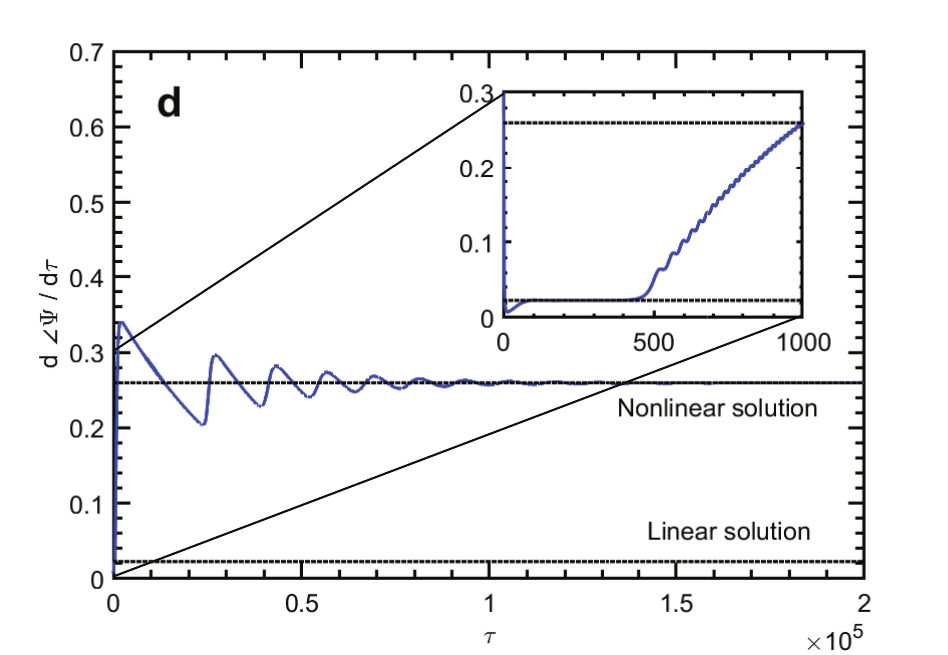}
            \put(50, 0){
                \colorbox{white}{
                    \makebox[0pt][c]{ $\tau$}
                }
            } 
            \put(-1.9, 30){
                \colorbox{white}{
                    \makebox[0pt][r]{\rotatebox{90}{\scriptsize Frequency}}
                }
            }
        \end{overpic}
    }
    
    \caption{Intensity and frequency for $\psi_1$ and $\psi_2$ with $\epsilon = 0.001 \kappa$, obtained through numerical integration of main text Eqs.~(4), (18), and (19) in normalized time $\tau = t \kappa$. The other parameters are $\gamma_1 = 0$, $f_0 = 0.9 \kappa$, and $\gamma_2 = 0.1 \kappa$. A fast gain relaxation rate of $\gamma_{||}/\kappa = 200$ is used for (a)--(b), and a much slower rate of $\gamma_{||}/\kappa = 5 \times 10^{-5}$ is used for (c)--(d). In either case, the intensities of $\psi_1$ and $\psi_2$ are slightly asymmetric due to the small perturbation, but their frequencies are equal. In both (b) and (d), the system briefly follows the linear solution prior to the gain being saturated. After a turn-on transient, it eventually reaches the steady-state solution shown in Section~III.}
    
    \label{fig:transients}
\end{figure*}

We examine gain relaxation rates varying over six orders of magnitude. As shown in Figure~\ref{fig:transients}(b) and (d), the frequency of the lasing mode briefly follows the linear solution prior to experiencing a turn-on transient, which strongly depends on the gain relaxation time.

After these transients, the gain and loss of each resonator saturate, and an equilibrium is reached corresponding to the same steady-state nonlinear solution as in Section~III of the main text.

This provides some insight into the results for the binary system in Ref.~\cite{Hodaei2017}. If a short, high-intensity pump pulse is used with a fast semiconductor gain medium, both lasing modes may temporarily turn on before the pump turns off, dropping them back below threshold before the steady-state solution is reached. Assuming the dynamics of the system are much faster than the repetition rate of the pump laser, the linear solution for the modal frequencies could then be observed.

\FloatBarrier

\bibliography{mybib_abbrev_noeprint_nov1_v2}%